\documentclass[floatfix, reprint, aps, prd,superscriptaddress]{revtex4-1}

\usepackage[english]{babel}
\usepackage[utf8]{inputenc}
\usepackage{amsmath,amssymb,amsfonts}
\usepackage{graphicx}

\usepackage{bm}
\usepackage{latexsym}
\usepackage{epsf}
\usepackage{rotating}
\usepackage{epsfig,graphics,rotate}
\usepackage{wrapfig}
\usepackage{array,hhline,dcolumn}%
\usepackage[normalem]{ulem}
\usepackage{hyperref}
\usepackage{placeins}
\usepackage{geometry}

\usepackage{lineno}

\begin{document}

\author{R. Acciarri}
\affiliation{Fermi National Accelerator Lab}

\author{C. Adams}
\email{corey.adams@yale.edu}
\affiliation{Yale University}

\author{J. Asaadi}
\affiliation{University of Texas at Arlington}

\author{B. Baller}
\affiliation{Fermi National Accelerator Lab}

\author{T. Bolton}
\affiliation{Kansas State University}

\author{C. Bromberg}
\affiliation{Michigan State University}

\author{F. Cavanna}
\affiliation{Fermi National Accelerator Lab}
\affiliation{Yale University}

\author{E. Church}
\affiliation{Pacific Northwest National Lab}

\author{D. Edmunds}
\affiliation{Michigan State University}

\author{A. Ereditato}
\affiliation{University of Bern}

\author{S. Farooq}
\affiliation{Kansas State University}

\author{R. S. Fitzpatrick}
\affiliation{University of Michigan}

\author{B. Fleming}
\affiliation{Yale University}


\author{A. Hackenburg}
\affiliation{Yale University}


\author{G. Horton-Smith}
\affiliation{Kansas State University}

\author{C. James}
\affiliation{Fermi National Accelerator Lab}

\author{K. Lang}
\affiliation{University of Texas at Austin}

\author{X. Luo}
\affiliation{Yale University}

\author{R. Mehdiyev}
\affiliation{University of Texas at Austin}

\author{B. Page}
\affiliation{Michigan State University}

\author{O. Palamara}
\affiliation{Fermi National Accelerator Lab}
\affiliation{Yale University}


\author{B. Rebel}
\affiliation{Fermi National Accelerator Lab}

\author{A. Schukraft}
\affiliation{Fermi National Accelerator Lab}

\author{G. Scanavini}
\affiliation{Fermi National Accelerator Lab}

\author{M. Soderberg}
\affiliation{Syracuse University}

\author{J. Spitz}
\affiliation{University of Michigan}

\author{A. M. Szelc}
\affiliation{Manchester University}

\author{M. Weber}
\affiliation{University of Bern}

\author{T. Yang}
\affiliation{Fermi National Accelerator Lab}

\author{G.P. Zeller}
\affiliation{Fermi National Accelerator Lab}

\collaboration{The ArgoNeuT Collaboration}
\noaffiliation

\begin{abstract}
The capabilities of liquid argon time projection chambers (LArTPCs) to reconstruct the spatial and calorimetric information of neutrino events have made them the detectors of choice in a number of experiments, specifically those looking to observe electron neutrino ($\nu_e$) appearance. The LArTPC promises excellent background rejection capabilities, especially in this ``golden'' channel for both short and long baseline neutrino oscillation experiments.  We present the first experimental observation of electron neutrinos and anti-neutrinos in the ArgoNeut LArTPC, in the energy range relevant to DUNE and the Fermilab Short Baseline Neutrino Program.  We have selected 37 electron candidate events and 274 gamma candidate events, and measured an 80\% purity of electrons based on a topological selection.   Additionally, we present a of separation of electrons from gammas using calorimetric energy deposition, demonstrating further separation of electrons from background gammas.

\end{abstract}

\title{First Observation of Low Energy Electron Neutrinos in a Liquid Argon Time Projection Chamber}

\date{\today}

\maketitle


\section{\label{sec:Introduction} Introduction}

The confirmation of neutrino oscillations \cite{Kamiokande, SNOPlus} has transformed the field of experimental neutrino physics. Subsequent measurements of neutrino oscillation parameters, mixing angles and mass splittings \cite{MINOS, Kamland, DayaBay, T2K, Nova} have pushed neutrino physics into the realm of precision measurements. Neutrino experiments are set to measure CP violation in the lepton sector \cite{CPviol,DUNE,HyperK} as well as the mass ordering of neutrinos \cite{MassH}.  The short-baseline neutrino anomalies, which may be consistent with the existence of an eV-scale sterile neutrinos \cite{Athanassopoulos:1995iw, An:2015nua, Aguilar-Arevalo:2013pmq, steriles}, may be resolved by neutrino experiments observing oscillations at short baselines (such as the Short Baseline Neutrino Program \cite{SBN}), along with the measurements of other experiments \cite{minosSteriles,TheIceCube:2016oqi,dayaBaySteriles}.

A particularly versatile method to probe neutrino physics in upcoming experiments is with the use of high-power neutrino beams. For many of the upcoming experiments above, there is a shared experimental signature:  the appearance of electron neutrinos ($\nu_e$) from an initially muon-neutrino ($\nu_{\mu}$) beam \cite{DUNE, SBN, Nova, microboone} using charged current (CC) interactions to identify the neutrino flavor.  For these experiments, the choice of initial beam energy spectrum and baseline allows the experiment to probe the relevant physics goals.

\begin{figure}[ht!]
  \includegraphics[width=\columnwidth]{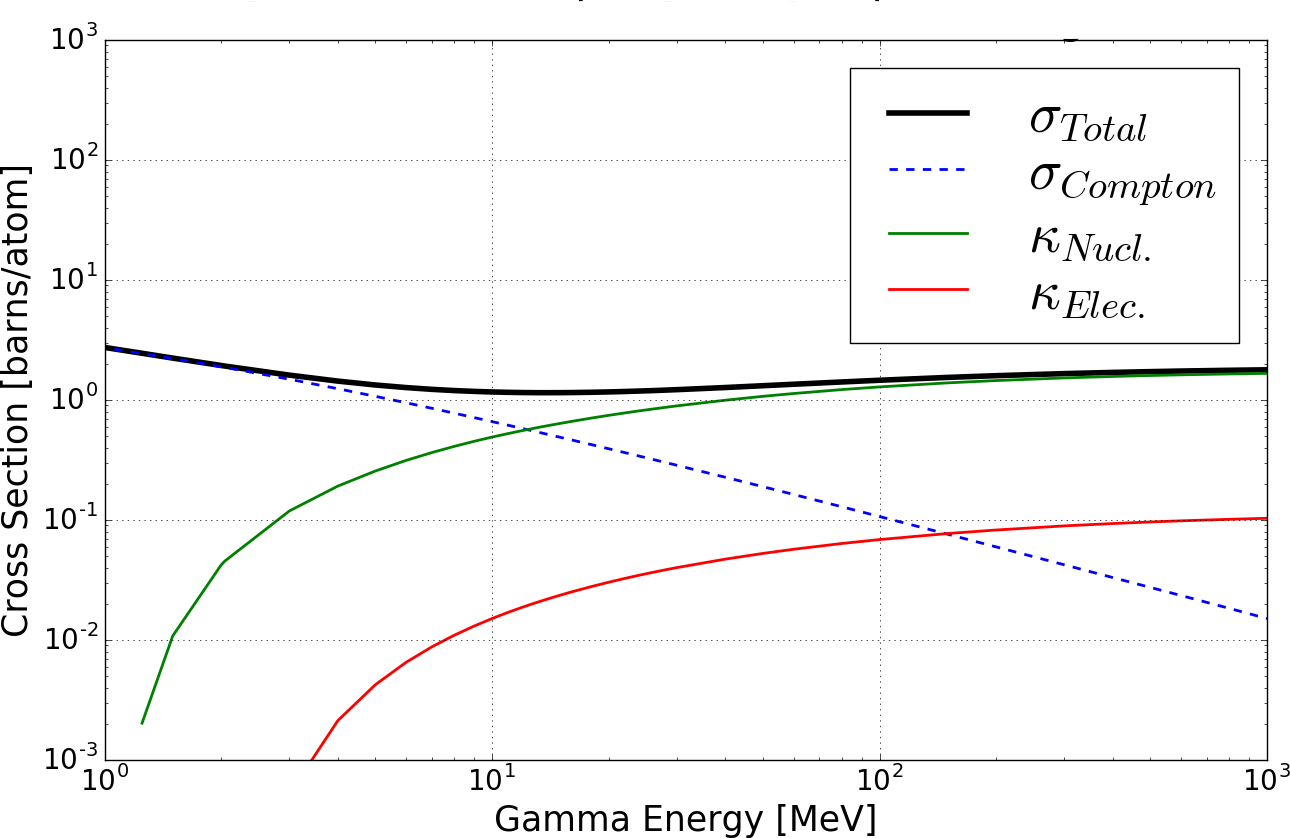}
  \caption{\label{fig:gamma_xsec} The cross section of high energy gammas on argon between 1 MeV and 1 GeV.  Most gammas produced by neutrino interactions relevant to DUNE \cite{DUNE} and the SBN Program \cite{SBN} are in this region.  Here, $\kappa$ refers to the pair production cross section for the nuclear field and electron field.  Pair production becomes the dominant cross section above 10 MeV.  Data obtained from the Xcom database \cite{Xcom}.}
\end{figure}

Historically, in many neutrino experiments, such as Cherenkov Imaging Detectors, a main background for CC $\nu_e$ events is neutral current interactions, which produce $\pi^0$ mesons equally for each neutrino flavor.  The $\pi^0$ mesons decay preferentially into pairs of energetic gammas. These gammas are hundreds of MeV for the Booster Neutrino Beam \cite{microboone,miniboone,SBN}, for example. A gamma, at typical neutrino beam energies, converts primarily through pair production - see Figure~\ref{fig:gamma_xsec}. Gammas can appear almost identical to electrons in most neutrino detector technologies, especially in the case that the two electromagnetic showers overlap or one of the gammas escapes the detector before interacting. A successful measurement of CP violation and resolution of short-baseline anomalies in neutrino physics requires high discrimination power between electron neutrinos and high energy gamma backgrounds.

The searches for CP-violation and a light sterile neutrino require high precision measurements of $\nu_e$ appearance.  However, the small interaction rate of neutrinos, coupled with the small oscillation amplitudes for oscillations, means the available sample of electron neutrinos in data is small compared to backgrounds.  This makes maximizing the detection of electron neutrinos  and suppression of background signals essential.  Liquid argon time projection chamber (LArTPC) \cite{Rubbia, Willis:1974gi} technology provides excellent electron neutrino detection and electron/gamma separation. The primary method of discrimination between electrons and gammas exploits the radiation length ($X_0 = 14$ cm) in argon, which is large compared to the excellent spatial resolution of TPCs. This means that a gamma can leave a visible gap between its origin and the place in the TPC where it interacts.  For an electron originating from a CC $\nu_e$ interaction no such gap will be present.  This paper applies this topological selection to identify a pure sample of low energy electron neutrino events in a liquid argon time projection chamber.

High energy gammas can, in some cases, interact at a sufficiently short distance from the neutrino's interaction vertex such that the gap from the vertex is not visible.  Further, the hadronic activity at the neutrino interaction vertex could be invisible in the TPC data, either because it consists of only neutral particles or because the particles are below detection threshold.  Without the presence of hadronic activity to distinguish the neutrino interaction vertex, it is not possible to observe a gap.  In these cases, a second method of electron/gamma discrimination is possible which uses calorimetry at the start of the electromagnetic (EM) shower.  An electron produces ionization consistent with a single ionizing particle, whereas the electron/positron pair produced by a gamma conversion produces ionization consistent with two single ionizing particles.  The calorimetric discrimination of electrons from gammas through the measure of ionization at the beginning of the electromagnetic shower is frequently referred to as $dE/dx$ discrimination.  

In this paper, we present an analysis of electromagnetic shower events from the ArgoNeuT detector, described in Section~\ref{sec:argoneut_detector}.  We develop techniques to select a sample of data with electromagnetic shower content, which we manually scan to classify events as electron neutrino candidates or gamma candidates.  Section \ref{sec:Selection} describes the automated selection criteria used to produce the sample of electromagnetic shower events (approximately ~6000 candidate events selected from more than 4 million triggers), and Section~\ref{sec:topological_sep} describes the criteria for the manual selection.  After the manual selection, 37 electron candidate showers and 274 gamma showers are selected.  In Section~\ref{sec:Reco}, we describe the details of the electromagnetic shower reconstruction and comparison of the electromagnetic showers to single particle Monte Carlo simulation of electrons and photons.  More details about the shower reconstruction algorithms are available in Appendix~\ref{appendix:reconstruction}.  As a validation of the electromagnetic shower reconstruction, we find that the most probable value of ionization at the beginning of an electron-induced electromagnetic shower is 1.76 $\pm$ 0.02 MeV/cm, in agreement with the theoretical value.  Based upon comparison with the single particle Monte Carlo, the topological selection produced a sample of electron neutrino candidates that was 80\% $\pm$ 15\% pure.

Events in this analysis are classified with a manually scanning step, and so we do not attempt to calculate a selection efficiency or compare with a full beam Monte Carlo. We do compare basic properties of the electron candidate sample with the expected electron neutrino and anti-neutrino content of the NuMI beam in Section~\ref{sec:electrons}.  Finally, in Section~\ref{sec:dedx_sep} we provide a demonstration of the calorimetric separation of electrons and photons through the $dE/dx$ discrimination. Though this type of technique has been used in previous neutrino experiments \cite{nutev,minerva,icarus_sterile}, this work presents the first demonstration of the feasibility of this method for discriminating electrons and gammas originating from neutrino interactions in liquid argon.

\section{The ArgoNeuT Detector \label{sec:argoneut_detector}}
Neutrino interactions are detected in the ArgoNeuT detector through the observation of final state charged particles from the neutrino interaction.  The charged particles, including electrons, protons, muons, pions and kaons, ionize the argon atoms as they traverse the liquid argon of the TPC.  These ionization electrons are drifted by the application of an electric field to planes of sense wires, where the drift electrons produce signals on the wires through either induction or charge collection.  The wires as a function of time, when arrayed in a two dimensional image, produce high resolution images of interactions in the TPC such as those seen in Figures \ref{fig:photons} and \ref{fig:electrons}.

The ArgoNeuT detector \cite{ArgoNeuT_tech} ran in the NuMI (Neutrinos from the Main Injector) beamline at Fermilab, outside of Chicago, IL, for six months in 2009-2010. The ArgoNeuT TPC was housed in a double walled, super-insulated cylindrical cryostat containing approximately 550 L of argon.  The TPC had an active volume of 47 $ (w) \times $ 40 $ (h) \times $ 90 $ (l) $ cm$^3$ resulting in an active mass of 170 L of liquid argon.  The cathode plane, a G10 sheet with copper metalization on the inner surface, was biased with a voltage of -23.5 kV for a drift field of 500 V/cm and a drift velocity of $\sim$1.6 mm/$\mu$s.  The drift electric field is regulated with field-shaping strips of copper, 1~cm wide and spaced 1~cm apart, plated on to the interior dimensions of the TPC such that the strips were perpendicular to the drift direction. 

The detector was instrumented with two planes of 240 sense wires, spaced 4~mm between wires and 4~mm between each plane, sampled every 198 ns. The wires were mounted at -30$^\text{o}$ and +30$^\text{o}$ to vertical and a third, non-instrumented plane, placed between the active volume and the wire planes, acted as a shielding plane.  The ArgoNeuT detector was not instrumented with a light collection system, and so the scintillation light produced by particle interactions in liquid argon was not collected in the ArgoNeuT detector.

ArgoNeuT was trigged in coincidence with the NuMI beam spill signal, with a $\sim$215~$\mu$s delay and a total NuMI spill duration of 9.7~$\mu$s.  For comparison, the maximum drift time from cathode to anode is 295 $\mu$s.  ArgoNeuT was installed approximately 100 meters underground directly in front of the MINOS near detector \cite{MINOS_detector}, which has provided muon spectrometry for many ArgoNeut analyses \cite{TGMuon}.  For more details on the construction and operation of the ArgoNeuT detector see \cite{ArgoNeuT_tech}.

\begin{figure}[htbp]
  \centering
  \includegraphics[width=0.95\columnwidth]{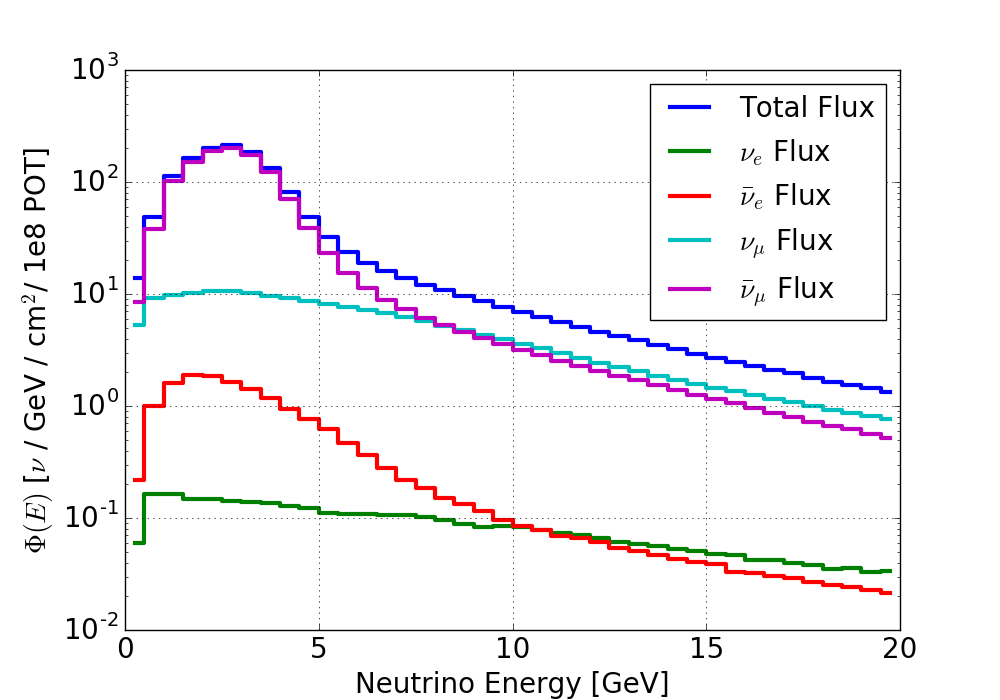}
  \caption{Neutrino flux at the ArgoNeuT detector in anti-neutrino mode.}
  \label{fig:argo_flux}
\end{figure}

The NuMI beamline \cite{NuMI} is the higher energy of the two neutrino beams produced at Fermilab. The beam is capable of running in neutrino and anti-neutrino modes, depending on the polarity of the magnetic field applied in the focusing magnetic horn system.  During the ArgoNeuT data taking, NuMI was running in the low energy mode, with the mean energy $\langle E_{\nu_\mu}\rangle = 9.6$ GeV, $\langle E_{\bar{\nu_\mu}} \rangle = 3.6$  ($\langle E_{\nu_\mu}\rangle = 4.3$ GeV in neutrino mode).  Although the beam consists mainly of muon neutrinos and anti-neutrinos there is a small ($\sim$2\%) contamination of electron neutrino and anti-neutrino events, with an energy spectrum shown in Figure~\ref{fig:argo_flux}.  This allows the study of electron neutrino interactions.   Data presented here were taken in both neutrino (8.5e18 protons on target (POT) ) and anti-neutrino mode (1.20e20 POT).

\section{\label{sec:Selection} Event Selection}

In ArgoNeuT, an event is defined as a readout window coincident with the trigger from the NuMI beam.  An event is much longer in time than a beam spill, however, to accommodate the drift time of electrons from the cathode to the anode.  Therefore an event consists of the collection of data from all 480 wires in the detector, read out over the 2400 ticks of digitization.  When the 240 waveforms of the wires of each plane are juxtaposed, and a color scheme is applied, an event can be visualized as seen in Figures~\ref{fig:photons} and \ref{fig:electrons}.  Due to the low interaction rate of neutrinos, events are typically empty (no significant ionization of any kind), and the next most common event contains externally produced particles, such as crossing muons from upstream interactions.  A small fraction of events contain neutrino interactions.  For this paper, for example, an `electron-like event' refers to the readout window of data that coincides with a candidate electron neutrino interaction in the TPC.

In order to demonstrate the calorimetric separation of electron-like events from gamma-like events, high purity samples of both electrons and gammas must be selected.  A sub-sample of the ArgoNeuT data set containing electromagnetic showers is isolated first through an automated procedure, and this sub-sample is used to select well defined electron and gamma events by visual scanning.

The selection criteria are determined from the ArgoNeuT Monte Carlo, using a GEANT-based simulation of interactions in the detector incorporated in the LArSoft package \cite{LArSoft}.  This Monte Carlo uses a Fluka simulation of the production of the flux \cite{fluka} to simulate the spectrum of neutrinos at the detector

\begin{figure}[htb]
\centering
\includegraphics[width=0.8\columnwidth]{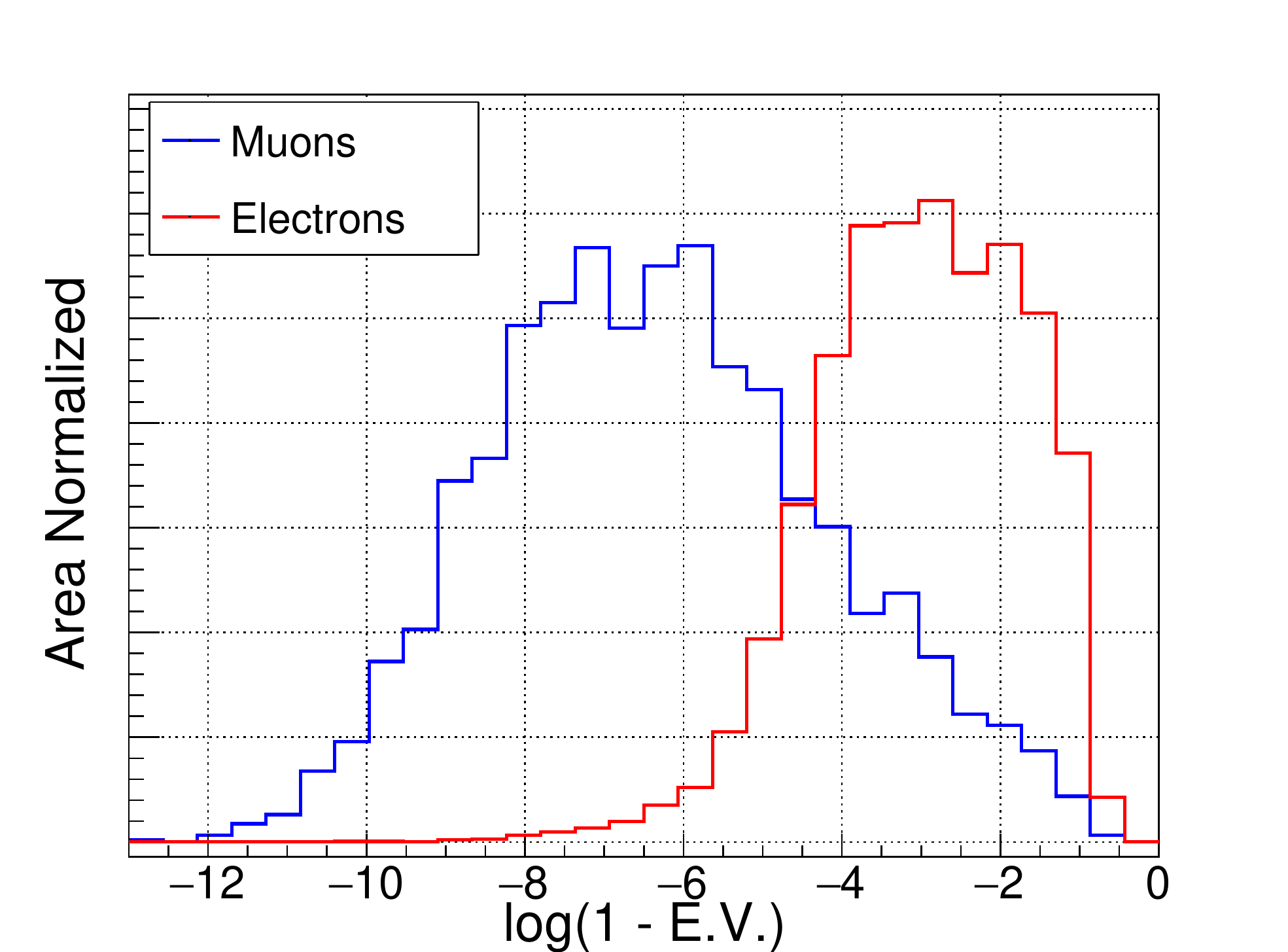}
\includegraphics[width=0.8\columnwidth]{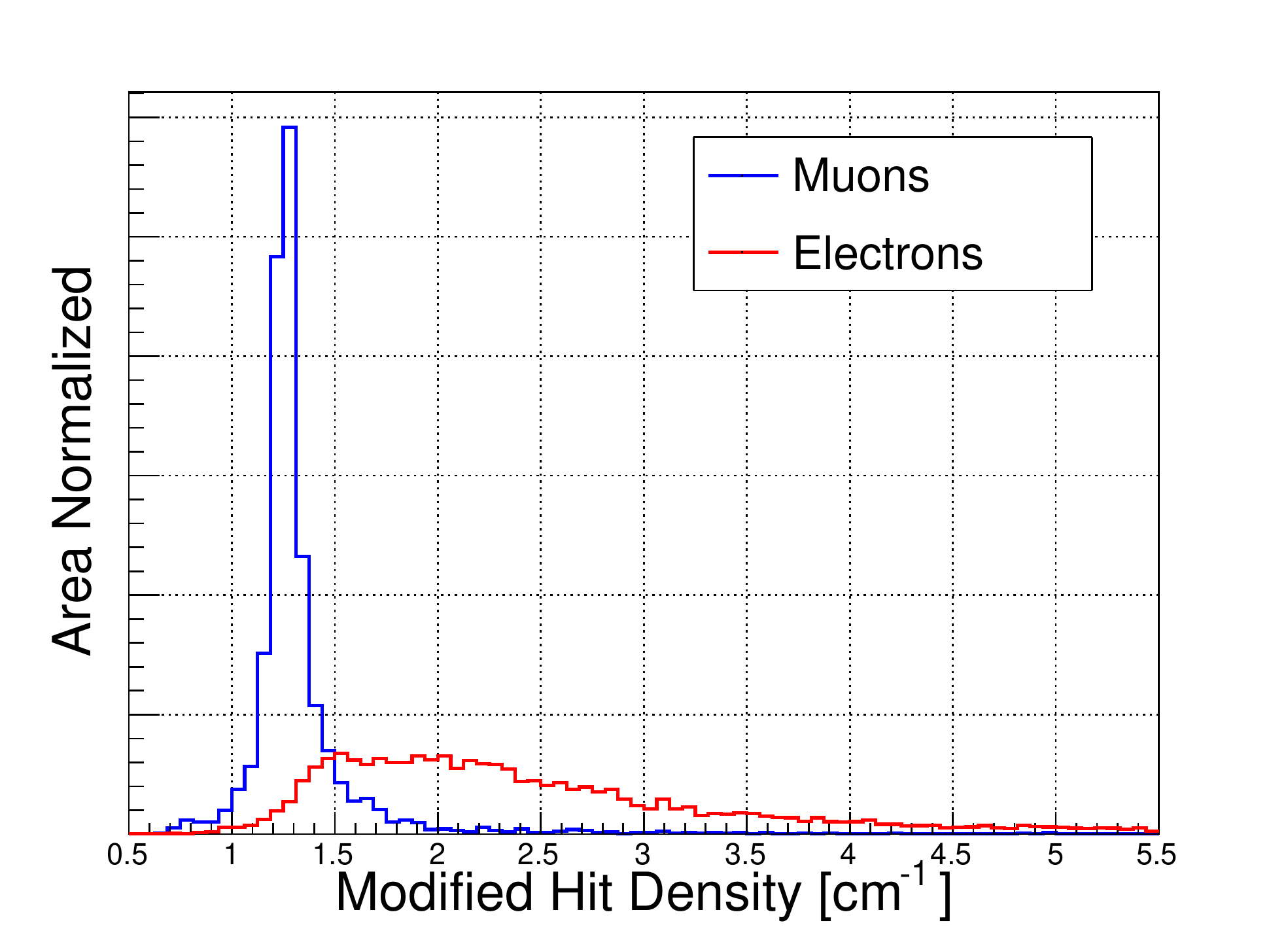}
\caption{\label{fig:separation} Principal Component Eigenvalue (top) and ``Modified hit density'' (bottom) calculated from Monte Carlo for single electron showers (red) and muon tracks (blue).}
\end{figure}

Selecting the sub-sample of electromagnetic showers is based on information from the 2-dimensional clusters of charge depositions (hits) in each wire plane. First, empty events and events with only track-like clusters are removed from the sample using an automated filter. This filter considers two-dimensional clusters of hits made with the LArSoft package \cite{LArSoft}, using an algorithm that is a combination of DBSCAN \cite{dbscan} and Hough Line Finding \cite{hough}, and calculates several parameters of these clusters to differentiate between track-like and shower-like clusters. 

The two most successful metrics in separating tracks and showers are the principal eigenvalue of a principal component analysis (PCA), and a direction corrected hit density of the cluster:
\begin{itemize}
  \item {\bf Principal Component Eigenvalue:} A principal component analysis (PCA) \cite{PCA} takes a collection of N-dimensional points and numerically finds the orthonormal coordinate system that best aligns to the data.  The goodness-of-fit metrics in the PCA analysis are the eigenvalues of the transformation matrix between the initial coordinate system and the best fit.  In this analysis, we use the 2D reconstructed charge depositions (hits) in the wire-time views of the collection plane TPC data and perform a principal component analysis on each cluster.  For track-like particles, which have strong directionality, the first eigenvalue of PCA is quite high, close to 1.  For shower-like clusters, the direction of the shower and it's transverse direction are less obviously separated, and the principal eigenvalue is lower than 1.  
  \item{\bf Direction Corrected Hit Density:}  A showering event is identified by significant activity in the TPC that is resolved away from the primary axis of the particle.  That is, a shower has many hits reconstructed as it travels through the TPC, whereas a track generally has one charge deposition detected per step through the TPC.  Measuring the hit density along a particle, defined as hits per unit distance, can thus discriminate between tracks and showers.  Since hits are only reconstructed on wires, the hit density is corrected to account for the fact that high angle tracks and showers (more parallel to the wires) have relatively fewer hits reconstructed.
 \end{itemize}

Figure~\ref{fig:separation} shows these separation parameters obtained using Monte Carlo simulations of single electrons as a model for electromagnetic showers, and single muons and protons as an archetype for tracks.  The Monte Carlo for this analysis is a GEANT4 based Monte Carlo through the LArSoft package, and we simulate single particles isotropically in the detector to determine separation properties \cite{geant, LArSoft}.  To select electromagnetic showers, a cut is made on the value of $log(1-E.V._{PCA}) > -5$ (see Figure~\ref{fig:separation} (top)).  $E.V._{PCA}$ is the first eigenvalue of the PCA analysis.  This corresponds to rejecting all clusters that have a principal eigenvalue greater than $\sim$0.999.  A second cut is made on the corrected hit density to reject track-like events.  Events with a corrected hit density greater than 1.5 hits per cm are kept (see Figure~\ref{fig:separation} (bottom)).  

An additional requirement is that a shower-like cluster in one plane should correspond to an analogous cluster in the second plane at the same drift time, measured by the time overlap of hits within the cluster. This removes spurious events tagged as showers due to wire noise or other sources in just one plane.

To remove events which resulted in a large amount of total charge, an additional set of criteria is applied using all of the hits in a single view in an event as a single cluster. These criteria remove high-multiplicity $\nu_{\mu}$ deep inelastic scatter events. 

This procedure resulted in a sample of ArgoNeuT events that contained an enhanced fraction of electromagnetic shower events, from which the final electron and gamma samples are identified through topological selection.  Table~\ref{tab:event_rates} shows the reduction of the ArgoNeuT data set by the automated filter.

\begin{table}[]
\centering
\caption{Reduction of ArgoNeuT data to the set of electromagnetic shower enhanced dataset}
\label{tab:event_rates}
\begin{tabular}{rcc}
\multicolumn{1}{l}{}                     & \multicolumn{1}{l}{\textbf{Neutrino}} & \multicolumn{1}{l}{\textbf{Anti-Neutrino}} \\ \cline{2-3} 
\multicolumn{1}{r|}{Beam Triggers}       & 445,812                                    &  \multicolumn{1}{|c|}{4,067,668}                     \\ \cline{2-3} 
\multicolumn{1}{r|}{Empty Event Filter } & 37,471                                     & \multicolumn{1}{|c|}{424,681}                     \\ \cline{2-3} 
\multicolumn{1}{r|}{Shower Selection}    & 765                                     & \multicolumn{1}{|c|}{5,692}                     \\  \cline{2-3}
\end{tabular}
\end{table}

\section{Topological Selection of Electrons and Gammas}
\label{sec:topological_sep}

\begin{figure}[htb!]
  \centering
  \includegraphics[width=0.8\columnwidth]{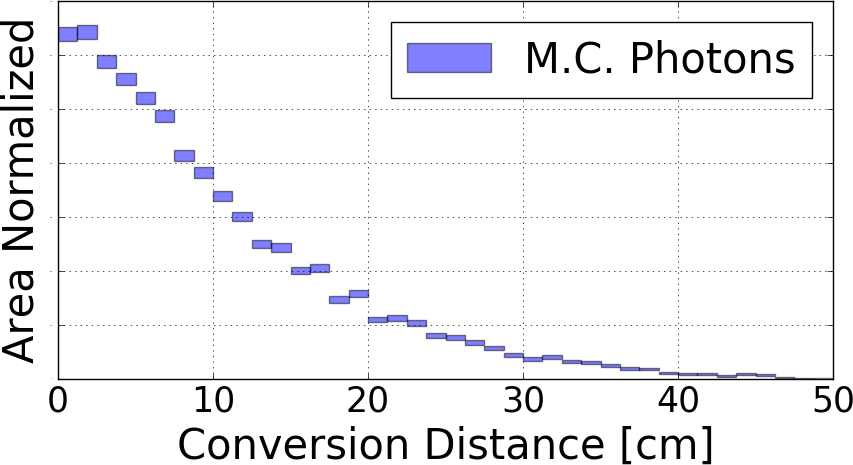}
  \caption{The conversion distance of each gamma in the Monte Carlo sample used for this analysis, which is about 7000 gammas in the energy range of several hundred MeV, as modeled by GEANT4 \cite{geant}. }
  \label{fig:photon_conversion_dist}
\end{figure}

When a gamma is produced in an interaction in argon, it will travel some distance, typically less than 50~cm (for a 500 MeV gamma), before it interacts and induces an electromagnetic shower.  Thus there is often a gap between the origin of the gamma and the start of the electromagnetic shower. If there is other activity in the detector at the location of the gamma production, the gap can be detected and the gamma can be classified. 

The simulated distribution of conversion distances for gammas in the energy range typical of the gammas used in this analysis is shown in Figure \ref{fig:photon_conversion_dist}.  There are gammas that convert very close to the generation point (here, 7\% of the gammas convert within a centimeter).  The definition of ``too close'' depends on the analysis being performed, however, there will always be a fraction of gammas for which a topological based cut is insufficient to tag them as gammas.  In the ArgoNeuT detector, the minimal resolution for a gamma gap is approximately one wire spacing (4~mm). In neutrino interactions with hadronic activity at the neutrino interaction vertex it is possible that other particles can obscure the start of an electromagnetic shower.  In this case, even gaps as large as a few centimeters can become unidentifiable.

\begin{figure}[ht]
\centering
\includegraphics[width=0.95\columnwidth]{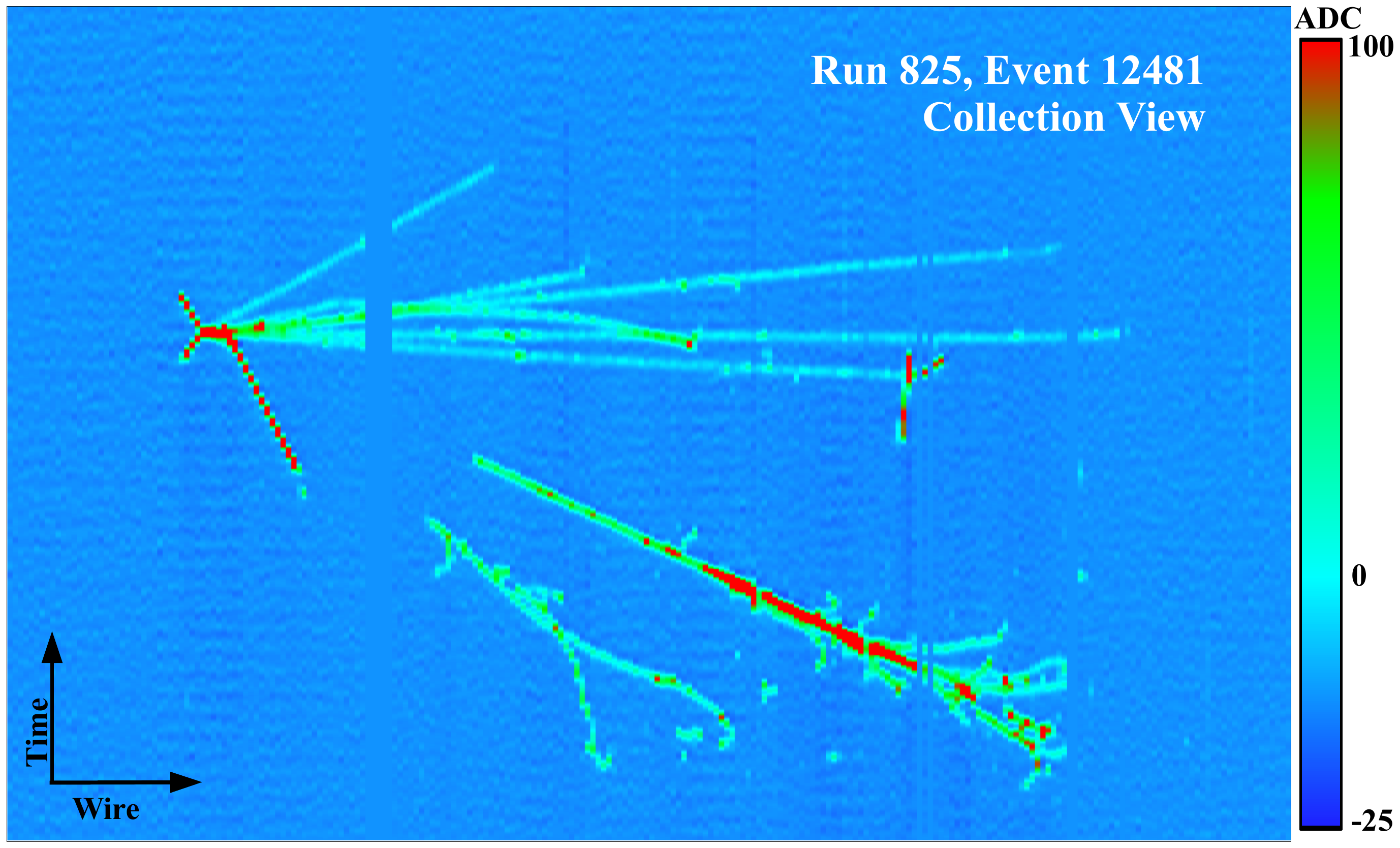}
\caption{\label{fig:photons} Example of an event with two gamma candidates in the ArgoNeuT data set.}
\end{figure}

We have chosen to define two types of topologies as gamma candidates, based on the observation of charged protons or pions at the neutrino interaction vertex: electromagnetic showers pointing back to charged particle activity at the displaced neutrino interaction vertex, implying hadronic activity, and $\pi^0$ candidate events.  In the second case, hadronic activity at the neutrino vertex is allowable but not required, and both electromagnetic showers are used in the analysis.  Example gamma interactions are shown in Figure~\ref{fig:photons}.  Gammas that we are unable to positively identify through only topological considerations - if, for example, the electromagnetic shower is the only activity in the detector - are removed from the data set entirely.

\begin{figure}[ht]
\centering
\includegraphics[width=0.95\columnwidth]{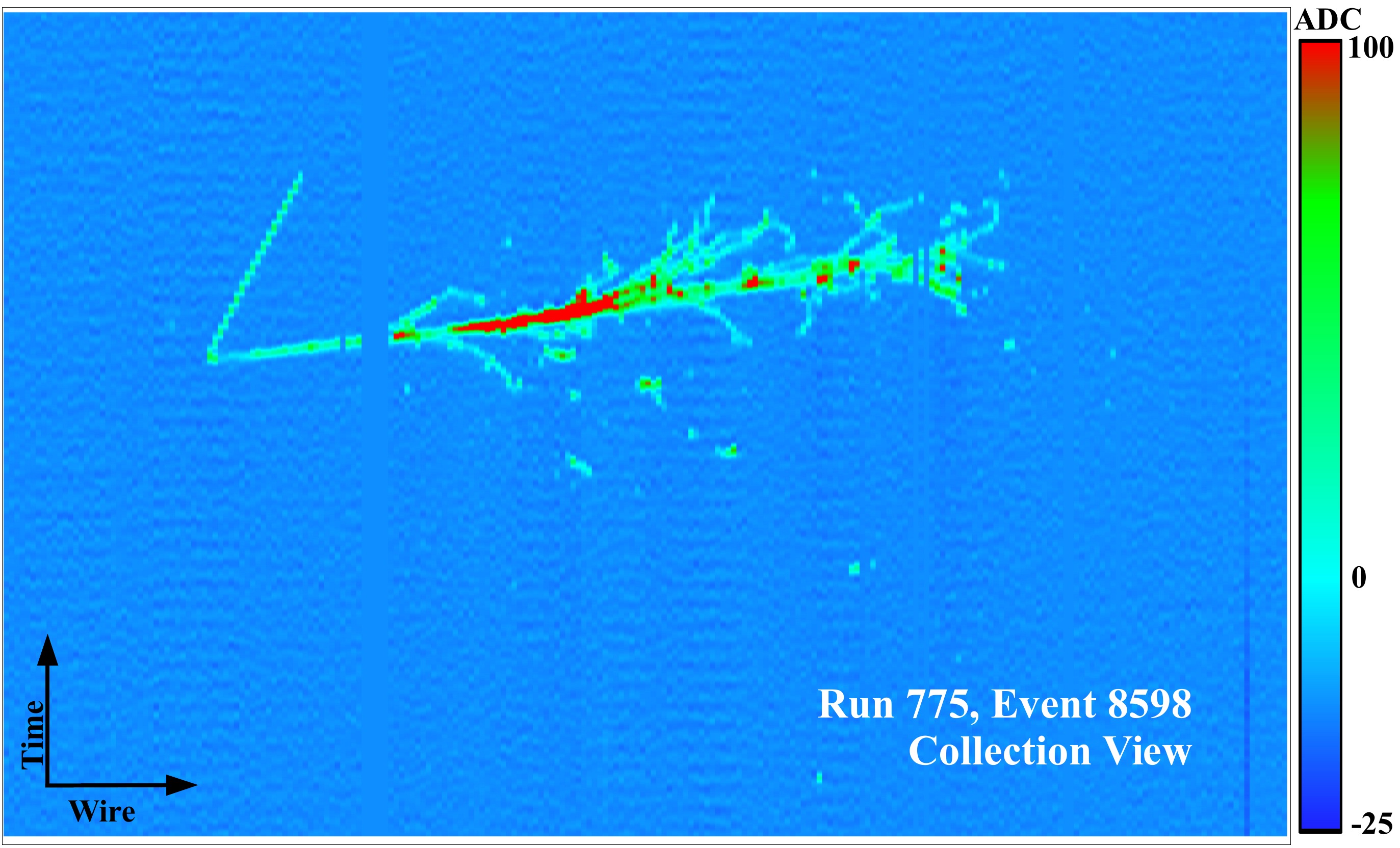}
\caption{\label{fig:electrons} Example of a $\nu_e$~CC event in the ArgoNeuT data set.  There is a region of dead wires that is located 20+ wires away from the neutrino interaction vertex, and is not considered a gap for selection purposes.}
\end{figure}

For a sample of electrons, this analysis targets electron neutrino events as the electron shower candidates.  To maximize purity, an electromagnetic shower is selected as an electron candidate only in events that also exhibited hadronic activity at the neutrino interaction vertex {\em without} the presence of a gap between the shower and other particles.  In addition, events with a track-like particle matched to a muon in the MINOS near detector are rejected.  This suppresses the $\nu_\mu$ charged current events in which the muon radiates significantly.  Of the events tagged as electromagnetic showers without a gap, 28\% were rejected because of a match to a track in MINOS.  An example of an electron candidate event is shown in Figure~\ref{fig:electrons}.  As a point of clarity, the ``gap'' in Figure~\ref{fig:electrons} is due to dead wires and not a region without ionization.  In this case, the cause of the dead wires is faulty electronics connection, and these electronics channels receive no signals from the TPC.  Since this ``gap'' is 20+ wires (8+~cm) from the neutrino interaction vertex, it does not impact the classification of this event.

The topological selection of events for this analysis is done manually, while the initial filter to select shower-like events is automated.  In total, 37 electron candidate showers and 274 gamma candidate showers are selected for the present analysis.  In the sample of gammas, 106 events are single gamma events while the rest have multiple gammas: predominantly two gamma $\pi^0$ candidates (69), 6 three-gamma events, and 3 four-gamma events.  This information is summarized in Table~\ref{tab:event_accounting}.

No electron events are fully contained in this analysis, as the ArgoNeuT detector is too small to contain GeV electron showers.  Some gamma events are contained, though not all, and it can't be determined on an event-by-event basis.  For the measurements presented later, the containment of the shower is not a critical parameter.  Instead, for calorimetric discrimination of electrons and gammas the behavior of the electromagnetic shower at the start of the shower activity is important.
  
\begin{table}[]
\centering
\caption{Summary of collected electromagnetic shower events, for the topologically selected electron and gamma categories.  By definition of the gamma topological selection, no gammas have hadronic overlaps.}
\label{tab:event_accounting}
\begin{tabular}{rl}
\multicolumn{2}{c}{\textbf{Electrons}}                         \\ \hline
\multicolumn{1}{r|}{Readout Events} & 37            \\ 
\multicolumn{1}{r|}{Hadronic Overlap} & 10           \\ \hline
\multicolumn{1}{r|}{\textbf{Total Electrons}}   & \textbf{37}  \\
\multicolumn{1}{l}{}                            &              \\
\multicolumn{2}{c}{\textbf{Gammas}}                            \\ \hline
\multicolumn{1}{r|}{Readout Events}             & 184          \\
\multicolumn{1}{r|}{Single Gamma}               & 106          \\
\multicolumn{1}{r|}{Two Gammas}                 & 69           \\
\multicolumn{1}{r|}{Three Gammas}               & 6            \\
\multicolumn{1}{r|}{Four Gammas}                & 3            \\ \hline
\multicolumn{1}{r|}{\textbf{Total Gammas}}      & \textbf{274}
\end{tabular}
\end{table}

\section{\label{sec:Reco} Electromagnetic Shower Reconstruction}

The selected electron and gamma candidates described in the previous section must be reconstructed to extract kinematic properties of the candidates’ associated neutrino interactions and secondary particles.  The first step in the reconstruction chain is to remove effects of electronics response and field response and remove electronics noise. This is done on a wire-by-wire basis using a Fast Fourier Transform based deconvolution kernel \cite{ArgoNeuT_tech}.  A signal peak finding algorithm is then used to find charge depositions on each wire, reconstructed as hits. The integral of the ADC count in each hit is used to calculate the charge $dQ$ using an (ADC$\times$Timetick)/Charge conversion constant. These constants are obtained using through-going muon events in a way analogous to \cite{TGMuon}, for every wire individually, on both collection and induction planes.  To determine the constants, all of the muon hits (from a separate analysis) on each wire are fit with a Gaussian-convolved Landau distribution.  The conversion constant is adjusted until the most probable value of the Landau distribution is 1.73 MeV/cm, the expected theoretical value \cite{TGMuon}.

The $dQ$ of each hit is corrected to account for the electron lifetime with an exponential formula $e^\frac{-t_{drift}}{\tau}$, where $\tau$ is the measured electron lifetime for each ArgoNeuT data run (typically 500 to 800 ms) and $t_{drift}$ is the time each charge deposition took to drift to the wires, calculated from its position in the TPC.  The drift time, $t_{drift}$, is known from synchronizing the ArgoNeuT readout window with the NuMI beam timing.  The uncertainty on the exact drift time, arising from the length of the NuMI beam spill, gives a small but negligible (1\% or less) uncertainty on the charge collected after to the lifetime correction.  The lifetime correction varies from a factor of $\sim$~1.75 (low purity runs at the cathode) to a typical correction of $\sim$~1.25. The measured charge deposition $dQ$ is also corrected for the recombination of electrons and ions as parameterized in \cite{Bruce}.

\begin{figure}[ht]
\centering
\includegraphics[width=0.95\columnwidth]{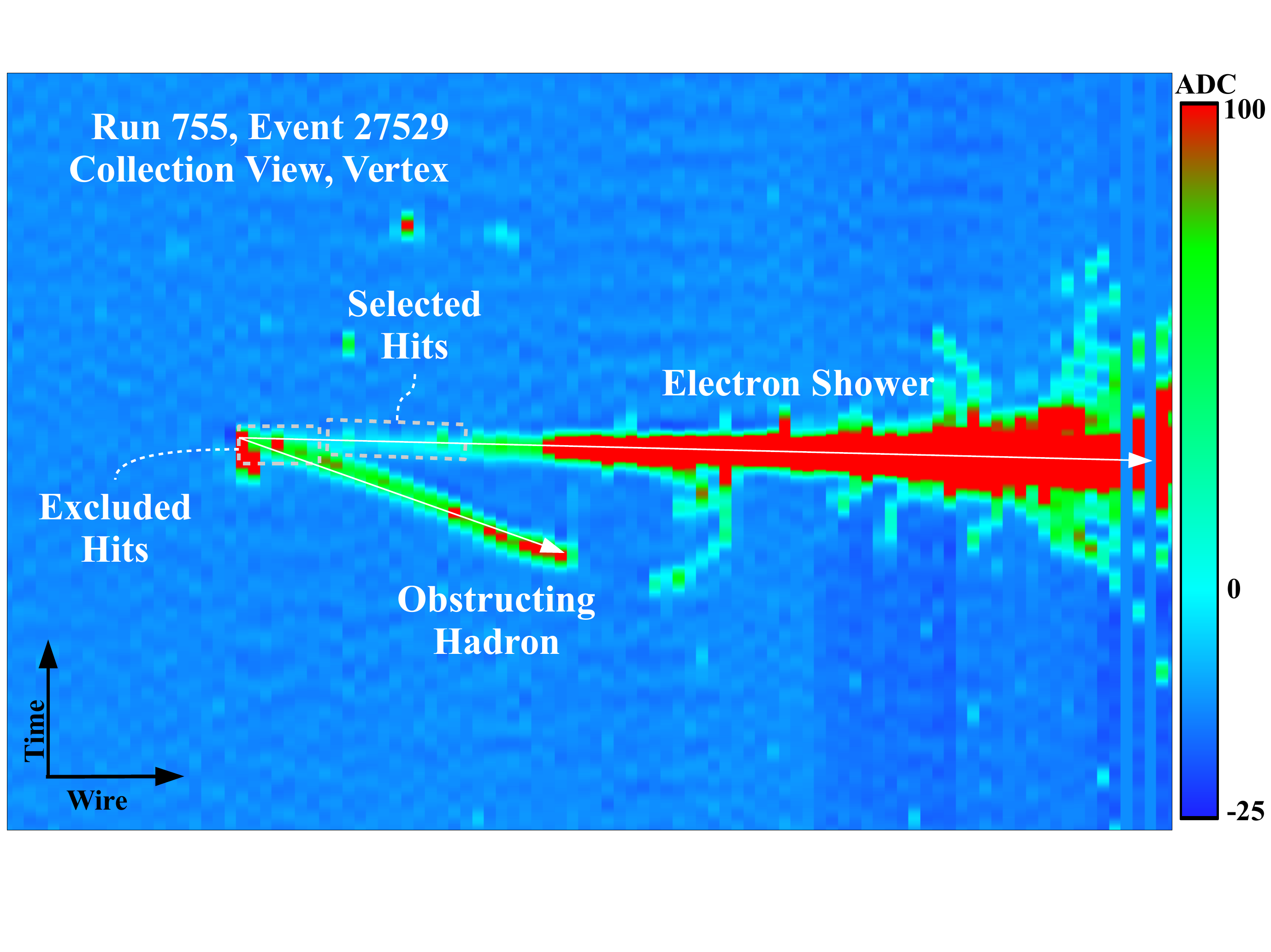}
\caption{\label{fig:hadronic-overlap} Example of a $\nu_e$~CC event in the ArgoNeuT data set with hadronic overlap.  For such events, the hits from the hadron are excluded manually from the analysis.}
\end{figure}

The hits for each candidate shower are reassembled into clusters using a manual scanning tool and fed into a shower-reconstruction algorithm. This allows the refinement of the start point and direction in each 2D plane for events with busy topologies.  In particular, for some events with overlapping protons and pions at the start of the shower, hits from the hadronic particle are manually excluded from the $dE/dx$ calculation.  This procedure is only done when the hadronic activity obscures a significant portion of the electromagnetic shower.  One quarter of the electron candidate sample had protons and pions obscuring the shower, in total.  An example of an electron candidate event with hadronic overlap is shown in Figure~\ref{fig:hadronic-overlap}.

The most important parameters that are computed in the reconstruction of electromagnetic showers are:
\begin{itemize}
  \item {\bf 3D Direction} - the direction of the shower, in space, is essential to the accurate calculation of the pitch of an electromagnetic shower as projected onto the wire planes.  For this analysis, the 3D direction is calculated at the start of the shower and not based on the full shower development, which can be affected by scattering of the primary showering particles.
  \item {\bf 3D Starting Point} - the 3D starting point is important to the electromagnetic shower reconstruction as it informs where the $dE/dx$ calculation should begin from.  This can be complicated by hadronic activity for showers close to a neutrino interaction vertex, though in this analysis the starting points have been verified manually.
  \item {\bf Deposited Energy} - the collection of all depositions of energy associated with electromagnetic shower are collected and summed to give an estimate of the amount of energy the initial particle left in the visible TPC.  Due to the small size of ArgoNeuT, electromagnetic showers are not well contained and the deposited energy is typically a fraction of the true energy of the incident particle.
\end{itemize}

In particular, the start point and direction are needed to measure the first several centimeters of the shower before the development of the electromagnetic cascade. Once the shower develops, the electron and gamma populations become significantly less distinguishable (see Appendix~\ref{appendix:dedx_calcs}). The details and validation of the electromagnetic shower reconstruction algorithms are available in Appendix~\ref{appendix:reconstruction}.

\begin{figure}[htb]
  \centering
  \includegraphics[width=0.8\columnwidth]{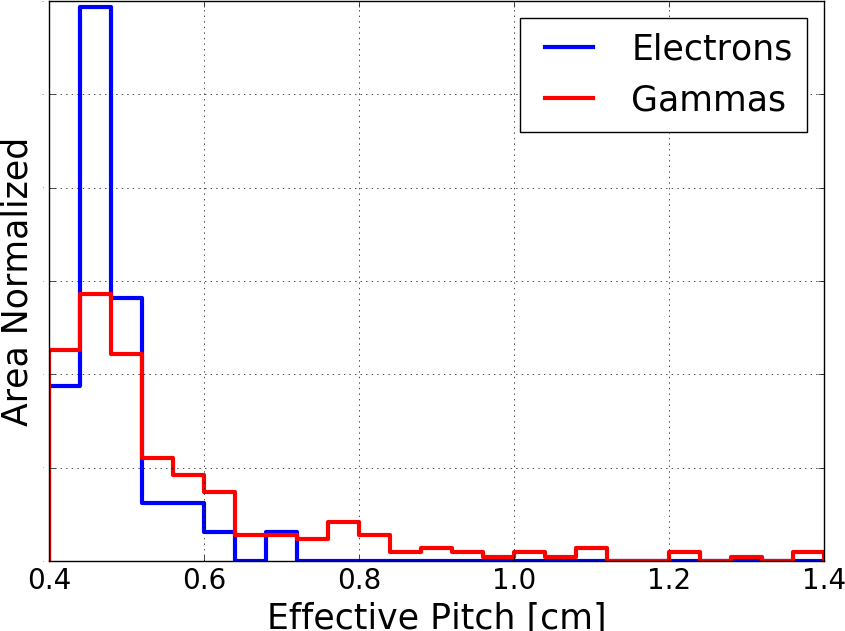}
  \caption{Effective pitch of hand-selected gammas and electrons in the ArgoNeuT dataset.}
  \label{fig:effective_pitch}
\end{figure}

For the calorimetric separation of electrons and gammas to succeed, the $dE/dx$ metric for each electromagnetic shower must be well reconstructed. As the charge depositions are measured discretely in 2D on single wires, in each of the wire planes we use the 3D axis of the shower to calculate an ``effective''  pitch ($dx$) between hits.  This effective pitch is, in other words, the real distance in the TPC that a particle travels between its two projections (hits) on adjacent wires. Figure \ref{fig:effective_pitch} shows the distributions of effective pitches for the electron and gamma samples.  The effective pitch is at least the wire spacing, which is 0.4 cm in ArgoNeuT.  The gamma distribution shows a slightly higher effective pitch, which is expected from Figure \ref{fig:geomety_dists} showing that the gammas are at slightly higher angles to the wire planes than the electron sample.  In the calculation of $dE/dx$, the effective pitch is used as the estimate of $dx$.

\begin{figure}[htb]
  \centering
  \includegraphics[width=0.95\columnwidth]{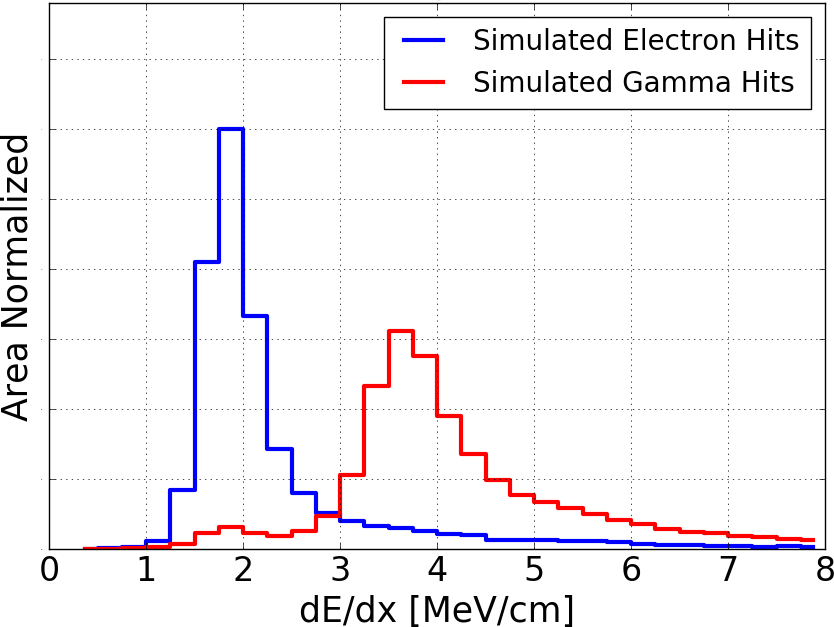}
  \caption{Distribution of $dE/dx$ for all hits at the start of the shower for the electron and gamma samples using Monte Carlo.}
  \label{fig:mc_landaus}
 \end{figure}

A valuable cross-check is the distribution of every $dE/dx$ deposition measured at the start of the shower.  Figure~\ref{fig:mc_landaus} shows the distributions for the Monte Carlo single particle simulation of both electrons and gammas.  The electron hits follow a Gaussian-convolved Landau distribution peaked at the $dE/dx$ value corresponding to one single ionizing particle. The gamma distribution peaks at a value corresponding to two single ionizing particles, but is more complicated due to the presence of gammas that Compton scatter instead of pair producing (seen at approximately 2 MeV/cm in Figure~\ref{fig:mc_landaus}).

\begin{figure}[htb]
  \centering
  \includegraphics[width=0.99\columnwidth]{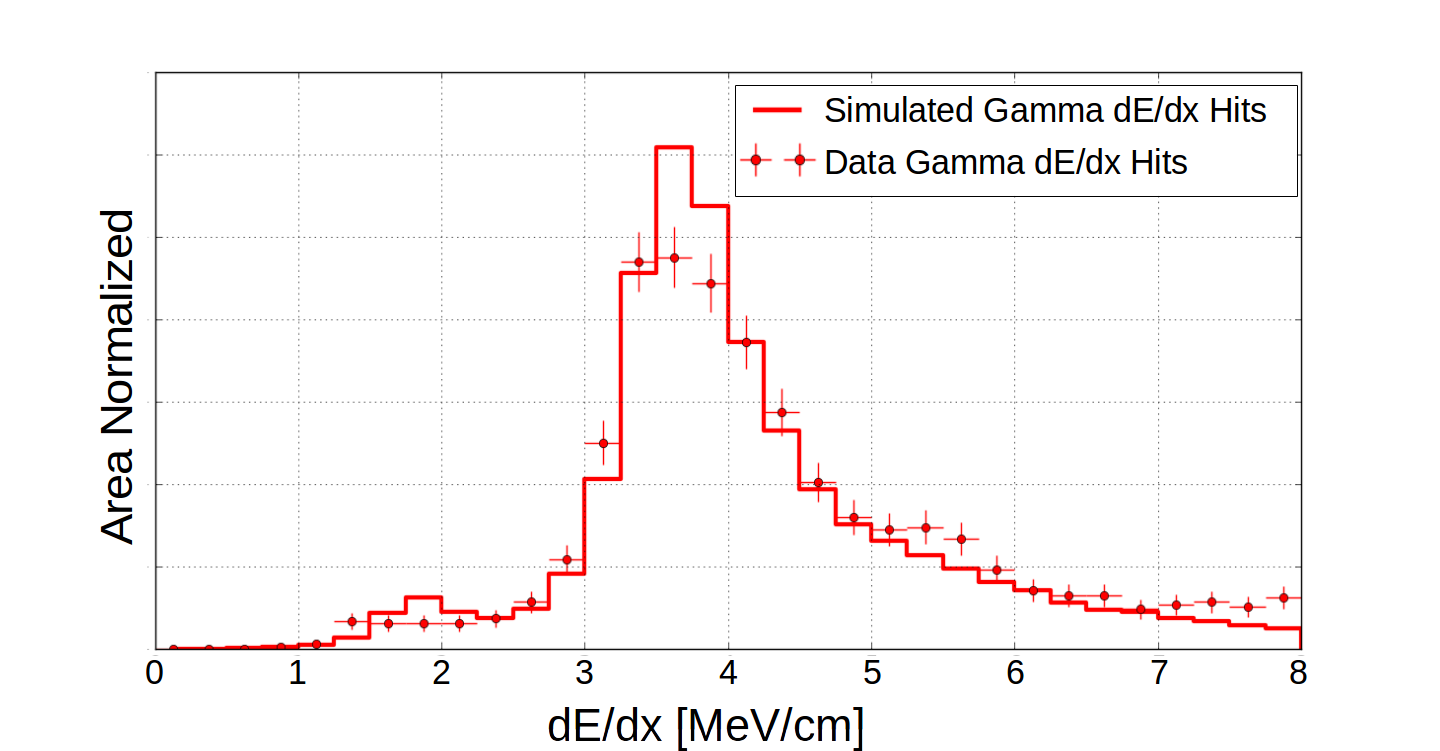}
  \caption{Distribution of $dE/dx$ for all hits at the start of the shower for the gamma sample.}
  \label{fig:photon_landau}
 \end{figure}

For the gamma sample, the comparison of data and simulation is shown in Figure~\ref{fig:photon_landau}.  Since the gamma sample is produced entirely by selecting showers that are displaced from the neutrino interaction vertex, the purity of the gamma sample is taken to be nearly 100\% in this analysis.  The Monte Carlo sample is a sample of single gammas produced with a Gaussian energy distribution, with the distribution tuned to best match data gamma distribution.  This tuning does not significantly affect the Monte Carlo distributions of $dE/dx$ except to change the relative proportions of the Compton and pair production populations.  Since the Monte Carlo is known to imperfectly model the data (single particle vs. neutrino induced gammas), a shape-only comparison is presented with area normalized distributions.  There is some disagreement in shape, particularly at the Compton population and the peak of the pair production population.  However, the data does appear to represent the Monte Carlo and the $\chi^2$/dof goodness of fit, based on the statistical uncertainties of the data distribution between 0 and 8 MeV/cm, is 1.4.

For the electron sample, we can not assume that the purity of the sample is 100\% based on topology alone.  As seen in Figure~\ref{fig:photon_conversion_dist}, a non-negligible amount of gammas will convert at a sufficiently short distance that they will be selected as electrons in a topological based cut.  Hadronic activity at the neutrino interaction vertex can also obscure the presence of a gap from a gamma.  Therefore, the distribution of electron-like $dE/dx$ hits analogous to Figure~\ref{fig:photon_landau} is expected to be modeled by a combination of electron and gamma showers in Monte Carlo.

\begin{figure}[htb]
  \centering
  \includegraphics[width=0.95\columnwidth]{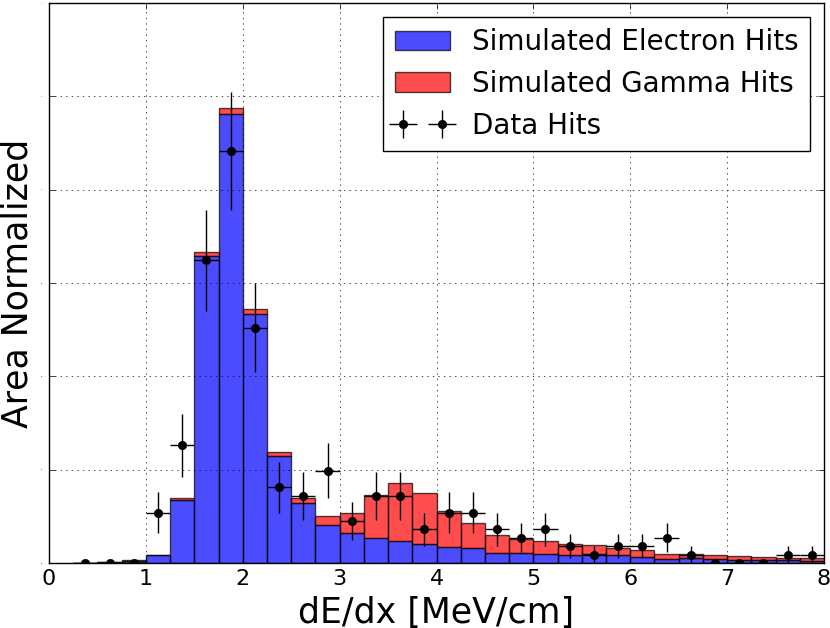}
  \caption{$dE/dx$ for all the hits from the electron candidate data sample, compared to a sample of Monte Carlo comprised of 80\% electrons and 20\% gamma.}
  \label{fig:electron_landau}
\end{figure}

The electron and gamma distributions from Figure~\ref{fig:mc_landaus} are used to fit the equivalent distribution of the electron-candidate data sample, using a linear combination of electron and gamma Monte Carlo such that the normalization of the total Monte Carlo distribution is normalized, consistent with the data distribution.  The $\chi^2$/dof is minimized between the (area normalized) data distribution and the combination of the electron and gamma distributions from Monte Carlo.  The best fit is shown in Figure~\ref{fig:electron_landau}.  The $\chi^2$/dof decreases from 2.78 with no gamma contamination to 1.02 when a gamma contamination is included at 20 $\pm$ 15\%, based on the data statistical uncertainties alone, and over the range of 0 to 8 MeV/cm.  The 15\% uncertainty is calculated as the width of the $\chi^2$/dof distribution at $\Delta \chi^2 = 1$.  This represents a direct measurement of the misidentification rate of the topological selection of electrons for this particular analysis, and demonstrates a method to measure this mis-ID rate in future electron neutrino searches in LArTPCs.

\begin{figure}[htb]
  \centering
  \includegraphics[width=0.95\columnwidth]{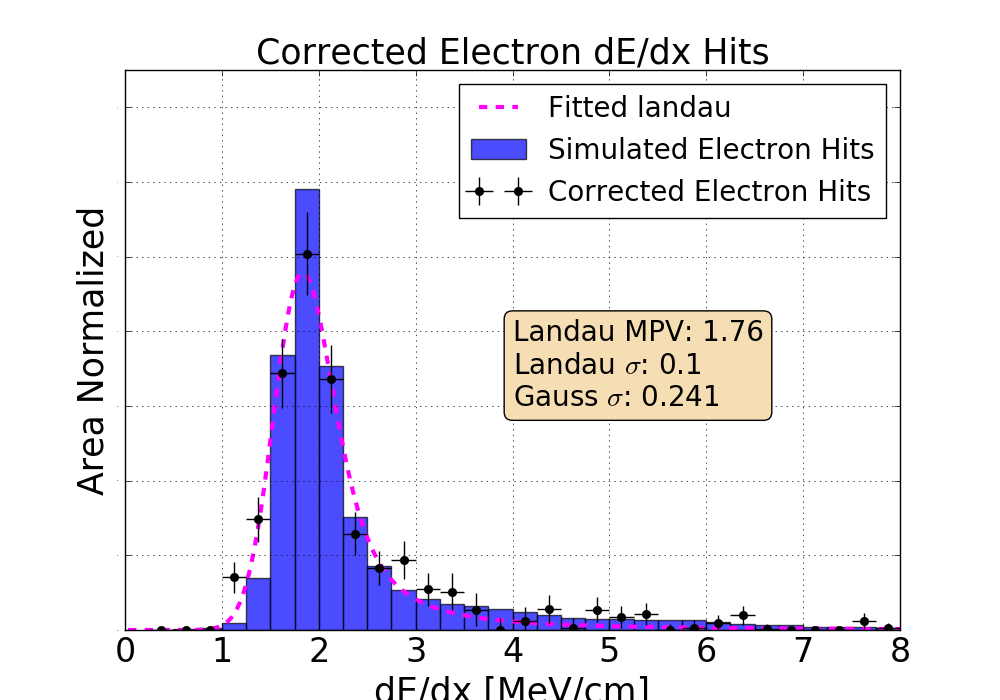}
  \includegraphics[width=0.95\columnwidth]{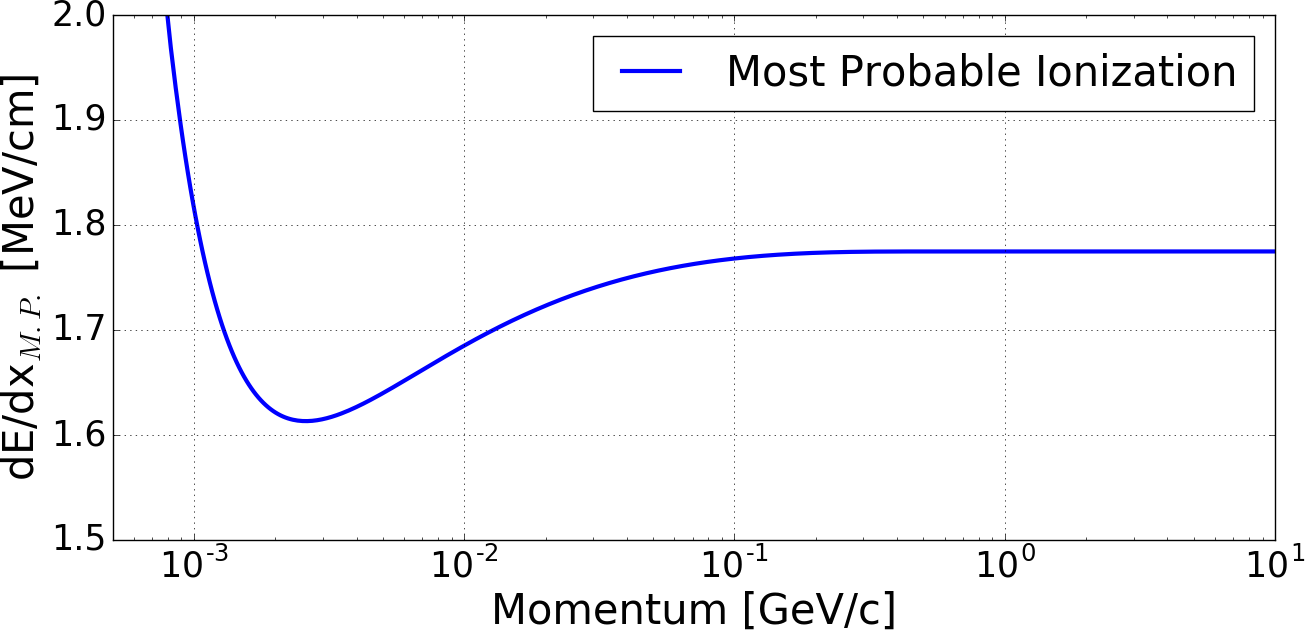}
  \caption{(Top) Background subtracted distribution of the hits at the start of the electron showers, with a fitted Gaussian-convolved Landau .  (Bottom) Most probable value of ionization as a function of momentum for electrons traversing liquid argon.}
  \label{fig:mpv_electrons}
 \end{figure}

As a final verification of the reconstruction, the measured distribution for the electron candidates is corrected by subtracting the gamma distribution from Figure \ref{fig:electron_landau}, scaled by the 20\% found above, using the gamma distribution from data.  This background subtracted distribution is fit with a Gaussian-convolved Landau distribution to determine the most probable value of charge deposition.  In particular, the most probable value of $dE/dx$ for electron-like hits is consistent with the theoretical values as shown in Figure~\ref{fig:mpv_electrons}. For electrons above 100 MeV/c, as this sample is, the theoretical expectation of the most probable ionization is 1.77 MeV/cm.  This is in good agreement with the fitted value of 1.76 $\pm$ 0.02 MeV/cm, where the error is computed from the covariance matrix of the fit parameters (again, statistical uncertainty only).

\section{\label{sec:electrons} Detection of Electron Neutrinos}
The sample of electron candidate events is expected to be exclusively from $\nu_e$ CC events.  As a validation,  we have studied the kinematic behavior of the electron-candidate sample.  Due to the small active volume of the ArgoNeuT detector, the electromagnetic showers are poorly contained and the initial electron energy is not a measurable quantity.  Instead, we measure the distribution of reconstructed {\em deposited} energy, and we compare it to a simulation of electron neutrino events.  The flux used to simulate the electron neutrino events is computed with a simulation of the NuMI beam with FLUKA \cite{fluka}. The electron neutrino and anti-neutrino flux for NuMI in anti-neutrino mode is shown in Figure~\ref{fig:argo_flux}($\langle E_{\bar{\nu}_e} \rangle$ = 4.3 GeV, $\langle E_{\nu_e} \rangle$ = 10.5).  The electron neutrino and anti-neutrino flux is predominately electron anti-neutrinos.

\begin{figure}[htbp]
  \centering
  \includegraphics[width=0.9\columnwidth]{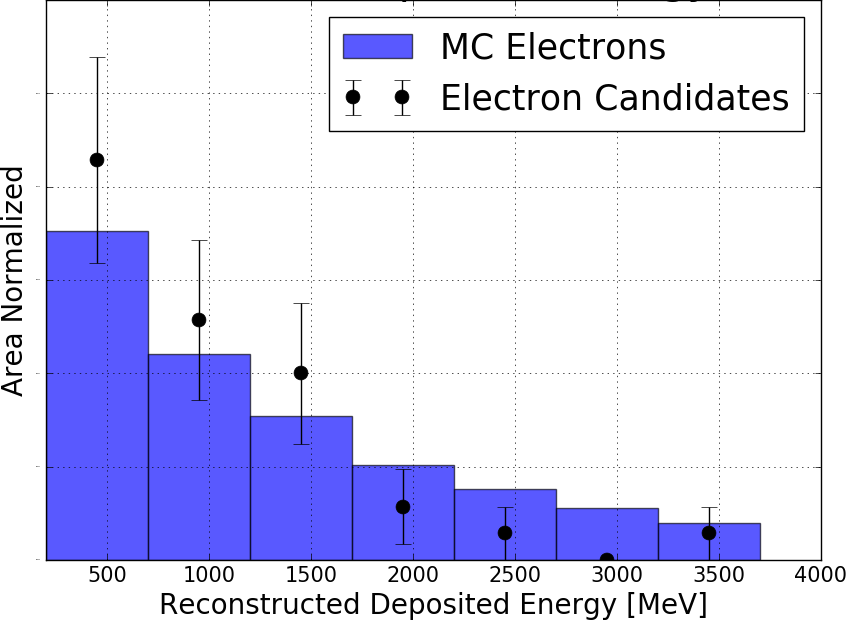}

  \vspace{0.3cm}

  \includegraphics[width=0.9\columnwidth]{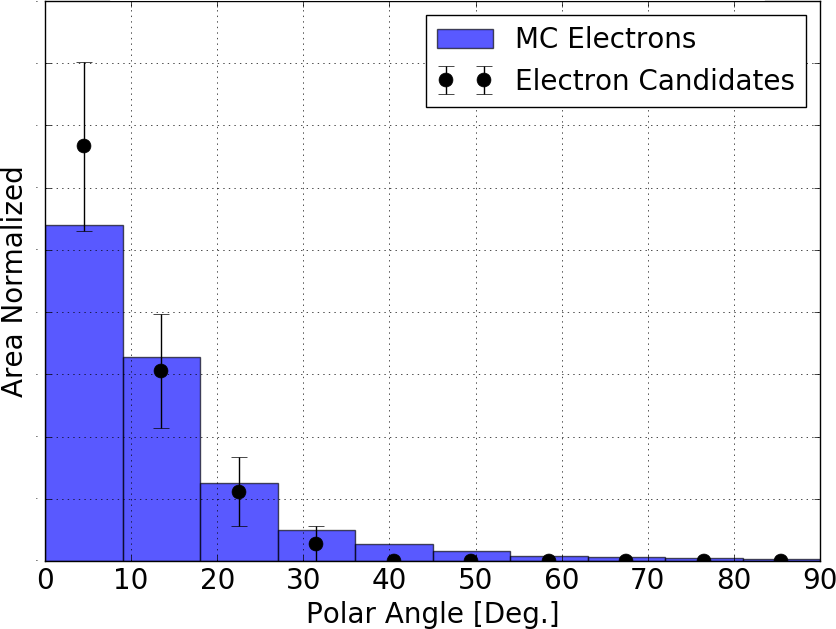}
  \caption{Kinematic distributions of deposited electron energy (top) and angle with respect to the beam (bottom). Error bars represent statistical uncertainty only.}
  \label{fig:electron_kinematics}
\end{figure}

Figure~\ref{fig:electron_kinematics} shows the kinematic distribution of the electron events' deposited energy and angle $\theta$, both calculated as described in Appendix~\ref{appendix:reconstruction}.  
Both the deposited energy and reconstructed angle are area normalized independently for both data and simulation.  In both distributions, the data has not been corrected to account for the 20\% contamination of gammas, whereas the simulation does not include any gamma contamination.  Despite this discrepancy, the distributions are presented as a demonstration that the electron candidate sample is well modeled by the Monte Carlo, despite the low statistics and other deficiencies.

Because the electron and gamma samples were selected with a manual method, we have not evaluated the Monte Carlo based efficiency of detecting these events.   Therefore, an absolute comparison of data and Monte Carlo is not presented here.  For the same reason, the measurement of the electron neutrino scattering cross section is also not presented.  In a subsequent publication, we will measure the selection efficiency with a fully automated selection and report the electron neutrino scattering cross section on argon.

\section{\label{sec:dedx_sep} $dE/dx$ Separation}

Once an electromagnetic shower has been identified and reconstructed, the information from the charge depositions at the start of the shower needs to be aggregated into a single dE/dx metric in order to separate electrons from gammas with calorimetry.

In the previous section, the conversion from $dQ/dx$ (the measured charge per unit centimeter), to $dE/dx$ (deposited energy per unit centimeter) is computed using a nonlinear model of the recombination of electrons and argon ions \cite{Bruce,birks}.  In considering the ionization at the start of a gamma induced shower where an electron and positron pair are present, we assume the ionization clouds of the two particles are sufficiently separated such that a non-linear model incorrectly inflates the $dE/dx$ from a $dQ/dx$, for higher values of $dQ/dx$.  Thus, the $dE/dx$ separation is computed using a minimally ionizing particle scale recombination correction for all charge depositions at the beginning of the shower in the electron and gamma samples.  While this is not applicable for highly ionizing fluctuations, it prevents an over estimation of the $dE/dx$ of gammas which artificially inflates the calorimetric separation power.  If the non-linear model of recombination was used, it would result in a photon peak at 20\% higher $dE/dx$, according to the parameterization in \cite{Bruce}.

For a given event there is not a statistically large sample of energy depositions to use for measuring a robust average $dE/dx$.  Given the Landau nature of the energy deposition fluctuations away from the most probable value, it is not surprising that an aggregate metric will tend towards higher energy depositions per centimeter than the most probable value.   For this analysis, when computing the $dE/dx$ separation metric for a shower all of the hits within a rectangle of 4 cm along the direction of the shower and 1 cm perpendicular to the shower are collected, and the median is computed.  Details about this choice of $dE/dx$ calculation are found in Appendix~\ref{appendix:dedx_calcs}.

\begin{figure*}[htb]
  \centering
  \includegraphics[width=0.95\textwidth]{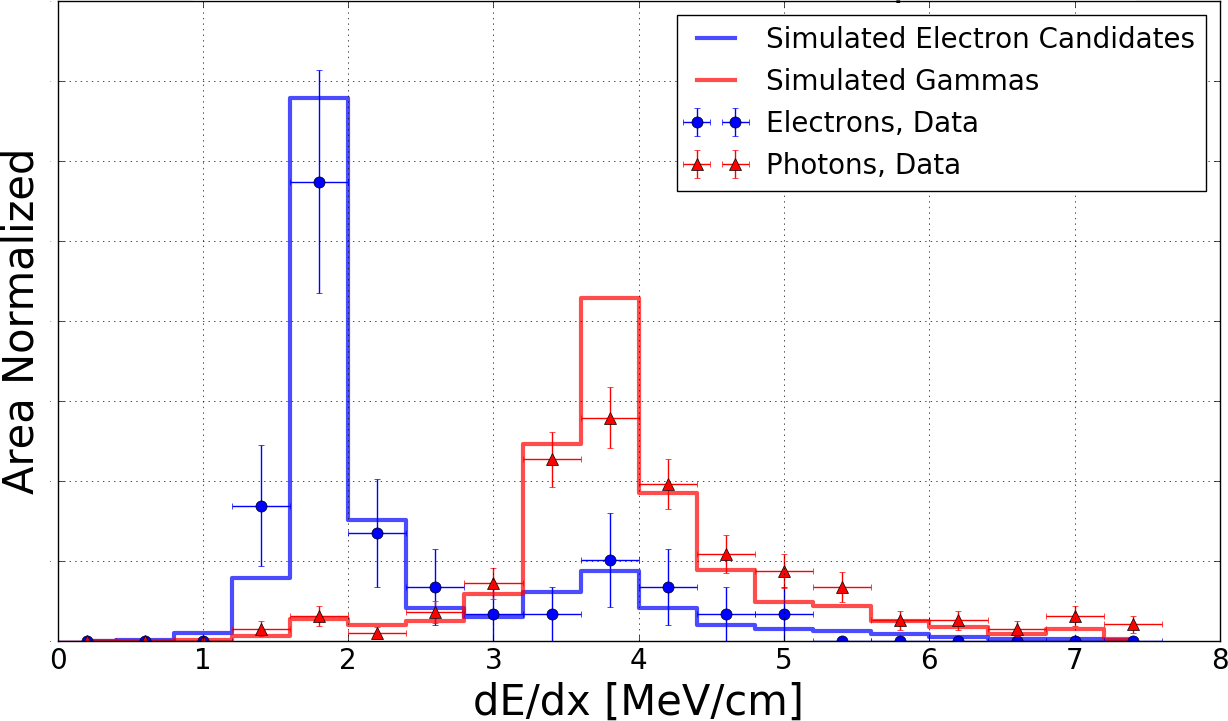}
  \caption{The $dE/dx$ distribution for electrons (blue) and gammas (red).  The solid blue curve, representing the simulation of electron $dE/dx$, includes a 20\% contaimination of gammas consistent with the results from Figure~\ref{fig:electron_landau}.}
  \label{fig:dEdx}
\end{figure*}

Results of the $dE/dx$ measurement of electrons and gammas are shown in Figure~\ref{fig:dEdx}. In contrast to Figures~\ref{fig:photon_landau} and \ref{fig:electron_landau}, Figure~\ref{fig:dEdx} represents the ability to discriminate between electrons and photons on an event-by-event basis.  This Figure represents the first demonstration of the calorimetric separation of electrons and gammas in a LArTPC using neutrino events.  Despite the low statistics of the ArgoNeuT experiment, the electron and gamma separation using calorimetry is clearly validated. For example, when a cut is made at 2.9 MeV/cm we find a 76 $\pm$ 7\% efficiency for selecting electron candidate events in data with a 7 $\pm$ 2\% contamination from the gamma sample. Here, the uncertainties on the efficiency are estimated with the Feldman-Cousins method \cite{FeldmanCousins} and are statistical only.  It must be noted, however, that the sample of electron candidates in this figure is not background subtracted.  The efficiency to select electrons with the same cut at 2.9 MeV/cm, estimated with the Monte Carlo, is 91\%.  This is consistent with the above measurement that 20 $\pm$ 15\% of the electron candidate sample, selected by topology only, is in fact gammas.  Lastly, the efficiency and purity of a $dE/dx$ selection metric will be impacted by the hit finding efficiency and wire spacing, and will vary amongst LArTPCs.

The value of the cut used above, 2.9 MeV/cm, is also somewhat arbitrary and must be determined uniquely for each analysis.  In this case, it is selected as the mid point between the two peaks of the distribution.  However, in an analysis targeting electron neutrinos the absolute normalization of the electron and gamma shower populations is crucial.  The desired purity of electrons must be balanced with the need to keep sufficient electron statistics.  An aggressive $dE/dx$ cut, at 2.5 MeV/cm, effectively rejects gammas but also can remove a significant amount of electrons (here it removes 30\% of electron candidate events in data, 13\% of Monte Carlo electrons).  Though this paper represents a demonstration of the calorimetric separation of electrons and gammas through $dE/dx$, it is strongly recommended to evaluate the precise values of the $dE/dx$ cut for future analyses.

\section{\label{sec:Conclusion} Conclusions}

We have analyzed a sample of neutrino events acquired by the ArgoNeuT detector and selected a sample of electron neutrino candidate interactions and gammas originating from neutral current and charged current muon-neutrino interactions. 

The high granularity of a LArTPC allows precision topological discrimination of gammas and electrons.  A purely topological cut produced a sample of electron neutrino events with an estimated 80 $\pm$ 15\% purity.  This is the first analysis to identify and reconstruct a sample of low energy electron neutrinos in a LArTPC.  The detection and characterization of these electron neutrino and anti-neutrino events is an essential step towards the success of large scale LArTPCs such as DUNE and the SBN Program.

Additionally, we have shown that a metric based on the $dE/dx$ deposition in the initial part of the shower is valid method of separating electron neutrino charged current events from gamma backgrounds, shown in Figure~\ref{fig:dEdx}. The full gamma background rejection capability of liquid argon detectors will be enhanced by adding to this a topological cut.  Further, full reconstruction of an event can improve gamma rejection.  For example, identification of two electromagnetic showers that reconstruct with an invariant mass consistent with the $\pi^0$ mass can remove both showers from the electron candidate sample, even if there is not a gap present and the $dE/dx$ cut fails. This work represents the first experimental proof of applying a calorimetric cut to separate electrons from gammas in a liquid argon detector using neutrino events.

One should note that the efficiency and misidentification rates presented here do not represent the full capability of liquid argon TPCs to discriminate gamma backgrounds from electron signals.  The final separation power of LArTPCs leverages multiple identification techniques, of which calorimetry is just one.  Further, the exact efficiencies and misidentification rates depend heavily on the energy spectrum of the electromagnetic showers: the Compton scattering gammas, a major source of impurity, appear predominately at energies below 200 MeV.

\section{Acknowledgements}

ArgoNeuT gratefully acknowledges the cooperation of the MINOS collaboration in providing data for use in this analysis. We would also like to acknowledge the support of Fermilab (Operated by Fermi Research Alliance, LLC under Contract No. DeAC0207CH11359 with the United States Department of Energy), the Department of Energy, and the National Science Foundation in the construction, operation and data analysis of ArgoNeuT. A.S. is supported by the Royal Society.  G.S. is supported by a Fermilab Neutrino Physics Center Fellowship.

\appendix 

\section{ Electromagnetic Shower Reconstruction}
\label{appendix:reconstruction}

\begin{figure}[htb]
  \centering
  \includegraphics[width=\columnwidth]{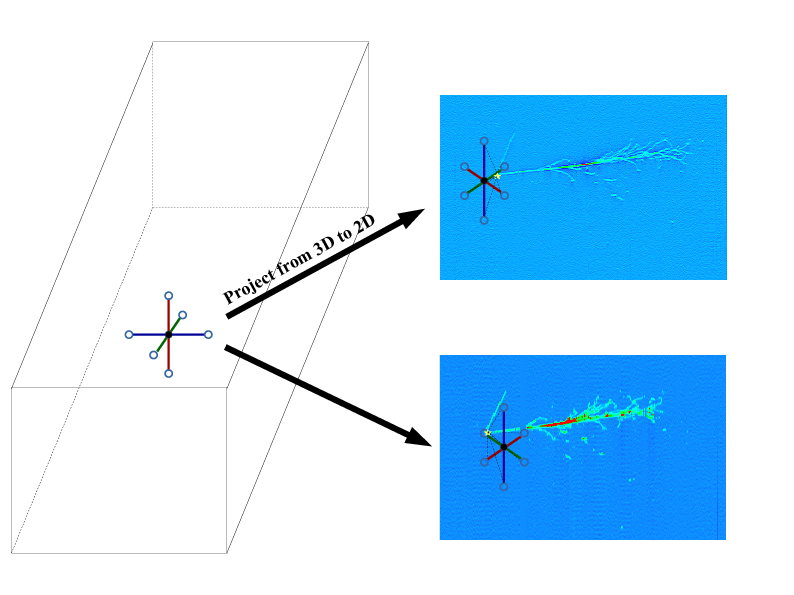}
  \caption{Diagram of the 3D start point algorithm.}
  \label{fig:iterative_start_point}
\end{figure}

The conventional coordinate system in LArTPC reconstruction algorithms assigns the Z direction to the direction of the beam, the Y direction as the vertical direction (bottom to top of the TPC), and the X direction in the drift direction such that the coordinate system is right handed.  The 3D start point is initially calculated from the intersection point of the wires where the two 2D start points are found and their position in the drift time coordinate.  The start point in 3D is improved using an iterative algorithm, and illustrated in Figure \ref{fig:iterative_start_point}.

\begin{figure}[htb]
   
   \includegraphics[width=0.95\columnwidth]{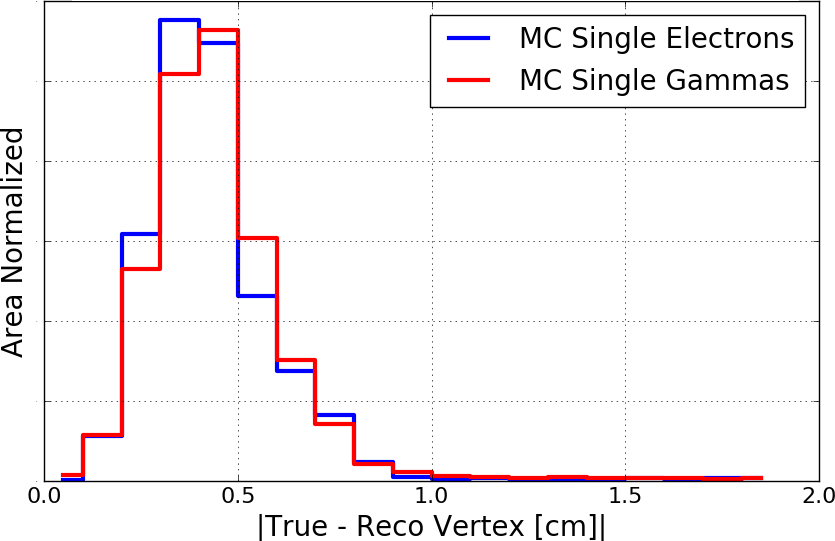}
   \includegraphics[width=0.95\columnwidth]{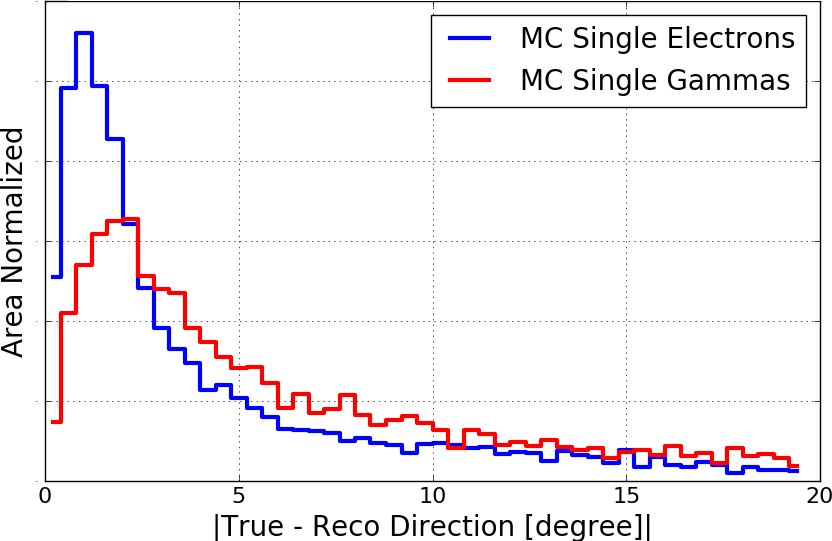}
   \caption{The calculated resolution of the 3D start point (top) and angular resolution (bottom) for single electromagnetic showers generated with the LArSoft package.  The angular resolution for gammas is slightly worse than for electrons because the gamma sample is at lower energy, and hence has fewer depositions (hits) in the TPC.}
   \label{fig:shower_reco}
\end{figure}

An initial guess, the point in black, is made for the start point based on the 2D start points (yellow stars in each plane).  The start point in 3D is projected into each plane, and the error in the 3D start point is the sum (over each plane) of the distance between the input 2D start point in each plane and the projection of the 3D point.  Six additional points, along the detector coordinates (in the $\pm$ x, y, and z directions), are also projected into each plane, and the error of each point is computed similarly (black dashed lines show the distance between projection and true start point).  The point with the smallest summed error is chosen as the improved 3D start point, and the algorithm makes an additional six guesses around it.  If the central point (in black) is chosen as the best fit point, the distance the other 6 points are offset from it is decreased and the algorithm repeats.  This procedure is repeated until the algorithm can no longer improve the accuracy of the 3D start point.  The initial offset from the central point for the 6 auxiliary points is 5 centimeters, and it decreases by 2\% for each successful iteration.  As seen in Figure~\ref{fig:shower_reco}, the 3D start point resolution is generally better than 1 cm.

\begin{figure}[htb]
  \centering
  \includegraphics[width=\columnwidth]{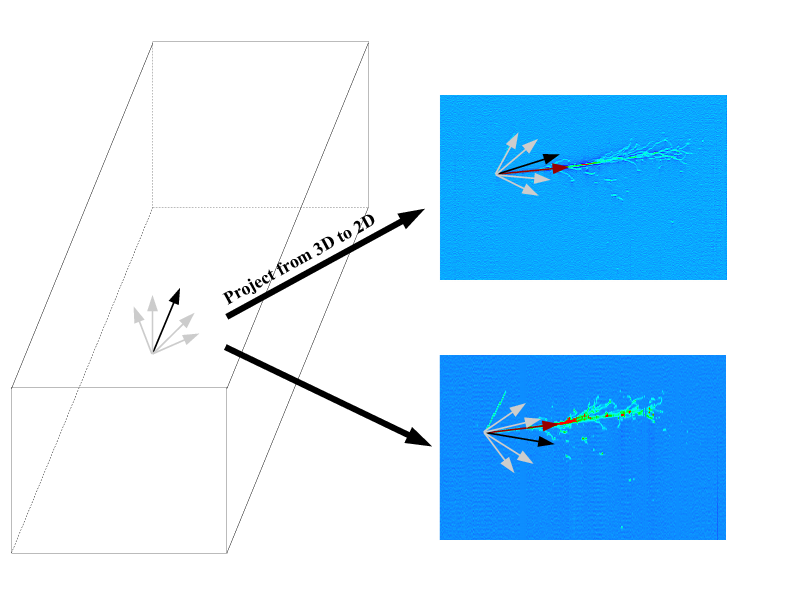}
  \caption{Diagram of the 3D start direction algorithm.}
  \label{fig:iterative_start_dir}
\end{figure}

Similar to the 3D start point, the 3D axis is computed using an iterative projection matching algorithm.  The standard TPC trigonometric formula is used to compute an approximate 3D axis based on the angle of each shower in the collection and induction plane:

\begin{align}
  \theta &= \text{arccos}\frac{m}{\sqrt{l^2 + m^2 + n^2}}, \\
  \phi &= \text{arctan}\left(\frac{n}{l}\right) 
\end{align}
where
\begin{align}
l &= sign(t_{end} - t_{start}), \\
m &= \frac{1}{2 \text{sin}(\alpha)}\left(\frac{1}{\Omega_0} - \frac{1}{\Omega_1}\right), \\
n &= \frac{1}{2 \text{cos}(\alpha)}\left(\frac{1}{\Omega_0} + \frac{1}{\Omega_1}\right).
\end{align}

Here, $\theta$ represents the polar angle in 3D with respect to the z axis (approximately the beam direction).  $\phi$ is the azimuthal angle in the x-z plane, with $\phi$ = 0 along the z axis, and  $\alpha$ is the angle of the wire planes with respect to the vertical direction, which in ArgoNeuT is 60 degrees.  $\Omega_0$ and $\Omega_1$ are the tangents of the 2D angles of the shower measured in each plane.  $t_{start}$ and $t_{end}$ are the start and end points of the cluster measured in drift time, such that $l$ is positive if the shower points away from the wires and negative if the shower points towards the wires.

\begin{figure}[htb]
   \centering

   \includegraphics[width=0.9\columnwidth]{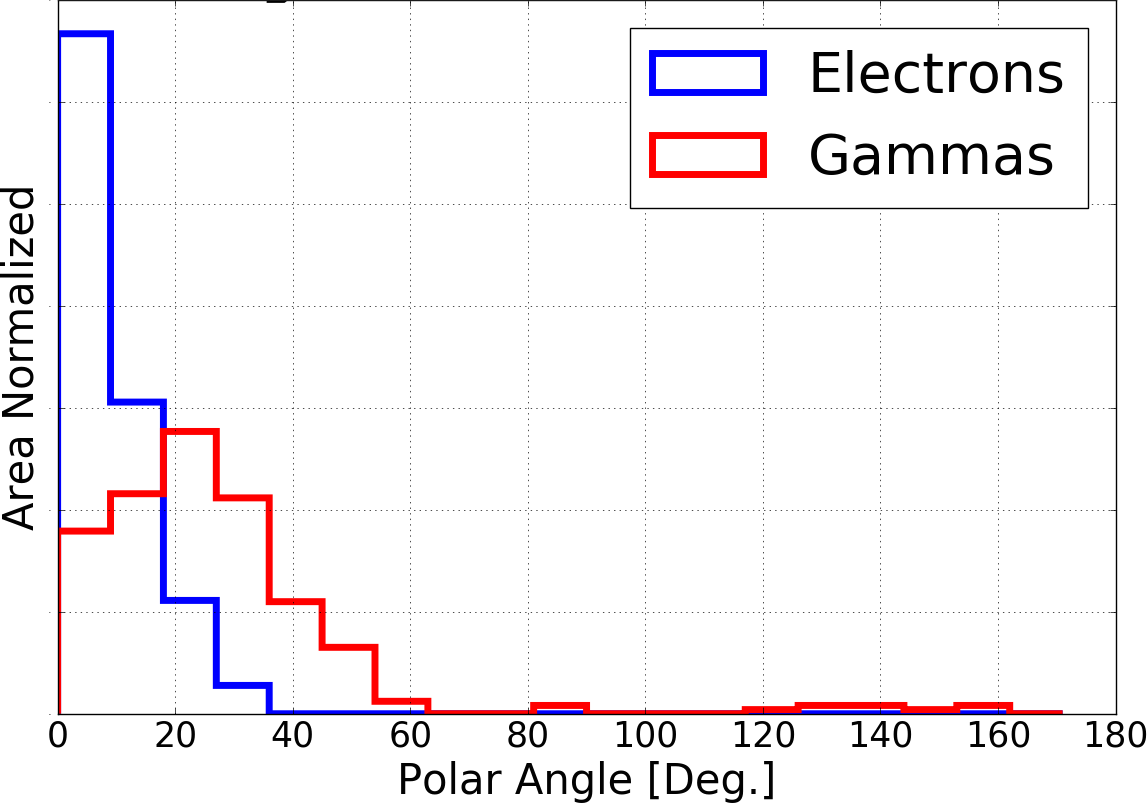}
   \caption{
   The distribution of the polar angle of events with respect to the Z direction (approximately the beam direction).  The electron sample is very forward going, and the gamma sample has a wider distribution of angles. }
   \label{fig:geomety_dists}
 \end{figure}

The reconstructed 3D axis is then projected into each plane, and the slope (in 2D) is compared against the slope of the electromagnetic showers in each plane.  Based upon the quality of the match between the projection and the 2D slopes, the 3D axis is adjusted until the best fit is obtained - see Figure~\ref{fig:iterative_start_dir}.  An initial guess, the arrow in black, is made for the start direction based on the 2D start directions (red arrows in each plane).  The start direction in 3D is projected into each plane, and the error in the 3D start direction is calculated.  An additional set of 3D directions (gray arrows) are also projected into each plane.  If the central direction (in black) is chosen as the best fit direction, the angular separation between it and the other (gray) directions is decreased and the algorithm repeats.  This procedure is repeated until the algorithm can no longer improve the accuracy of the 3D start direction.

The angular resolution for electromagnetic showers, shown in Figure~\ref{fig:shower_reco}, is generally quite good ( $< 5^\text{o}$) though there is a substantial tail.  However, for this analysis, the poor resolution in a few measurements of the 3D axis has a minimal effect on the $dE/dx$ calculation.  This is due to the fact that the majority of the events are forward going, as shown in Figure \ref{fig:geomety_dists}.  Therefore a moderate uncertainty in the 3D angle leads to only a small uncertainty in the effective wire pitch, described below, and a small uncertainty in $dE/dx$.


Since an electromagnetic shower is a combination of many single ionizing particles - electrons and positrons - and is not composed of highly ionizing stopping particles - i.e., protons - the measured charge on the sense wires in the peak of the showering activity is a sum of many minimally ionizing particles.  Therefore, to calculate the total energy deposited by an electromagnetic shower, each deposition collected is corrected by a recombination amount that is proportional to a minimally ionizing particle.  All of the energy depositions, once corrected, are summed into a final measure of the reconstructed, deposited energy.  


\section{ $dE/dx$ Calculation Methods}
\label{appendix:dedx_calcs}

While investigating the methods to convert a sample of hits (per shower) into a single variable, three promising $dE/dx$ metrics were developed:
\begin{enumerate}
  \item {\bf Outlier Removed Mean}: For every hit considered for each shower (within a certain distance from the start), the mean $dE/dx$ of the hits is calculated, as well as the RMS.  The hits that are outside of the mean $\pm$ the RMS are then rejected, and the mean of the remaining hits is recomputed and used.
  \item {\bf Median}: The same initial set of hits as above is used.  However, a median is calculated instead of rejecting outliers.  In particular, this method is robust against single high or low fluctuations.
  \item {\bf Lowest Moving Average}: For the same set of N initial hits, a moving 3 hit average is calculated.  For example, for N hits, the average is calculated of the hits (1,2,3), then the hits (2,3,4), etc. until the hits (N-2, N-1, N).  For all of these average values calculated, the lowest value is used as the $dE/dx$ measure.  This is designed to find regions where the start of the shower is behaving as a minimally ionizing particle for an extended period.
\end{enumerate}

\begin{figure*}[htb]
  \centering
  \includegraphics[width=0.32\textwidth]{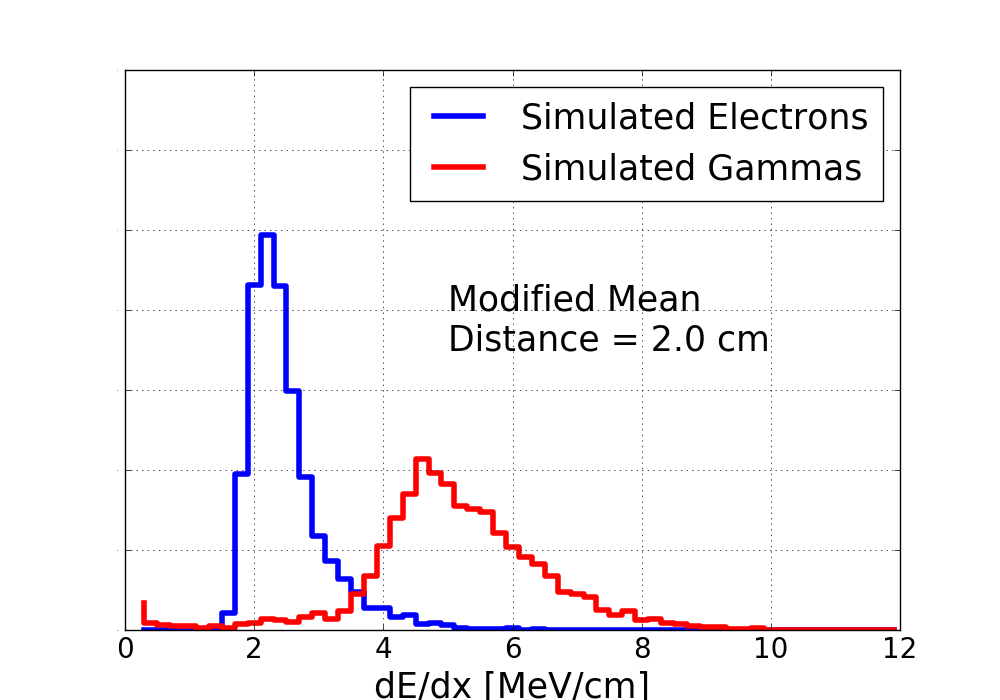}
  \includegraphics[width=0.32\textwidth]{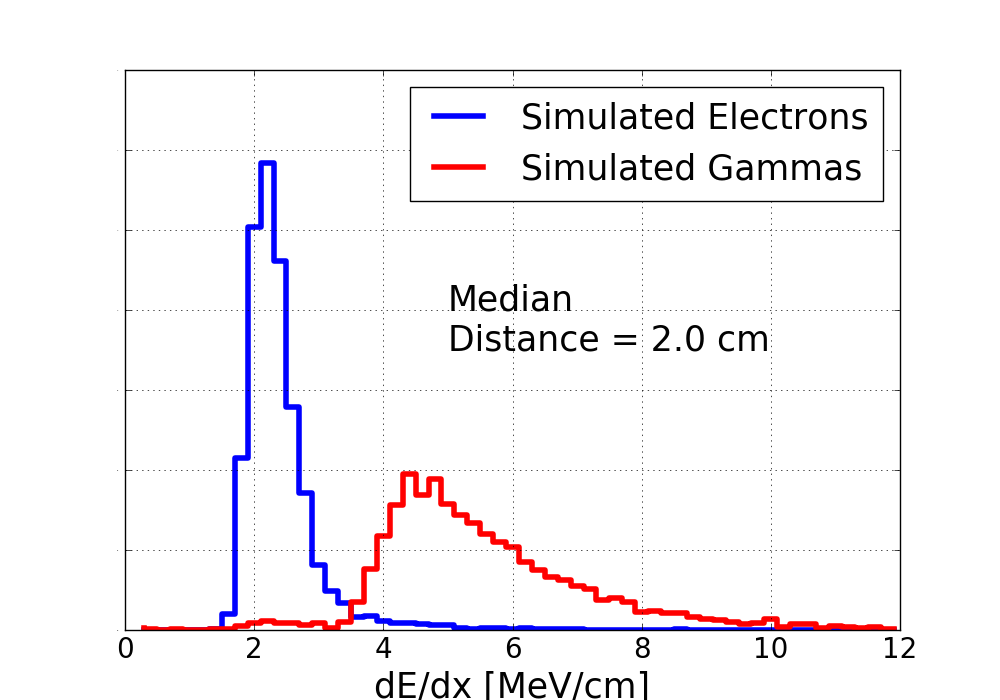}
  \includegraphics[width=0.32\textwidth]{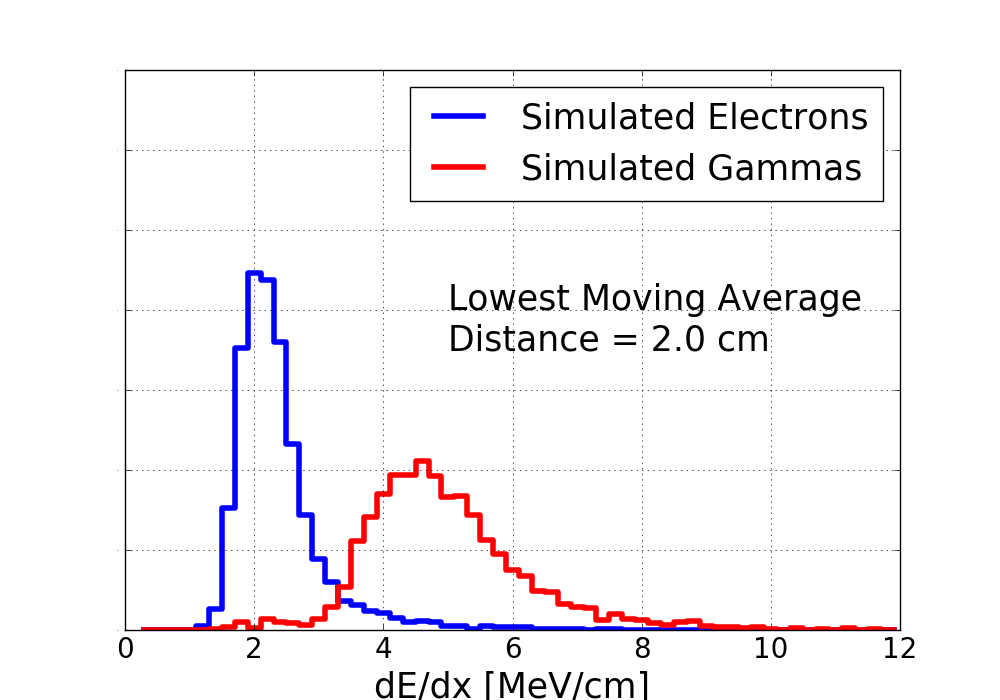}

  \includegraphics[width=0.32\textwidth]{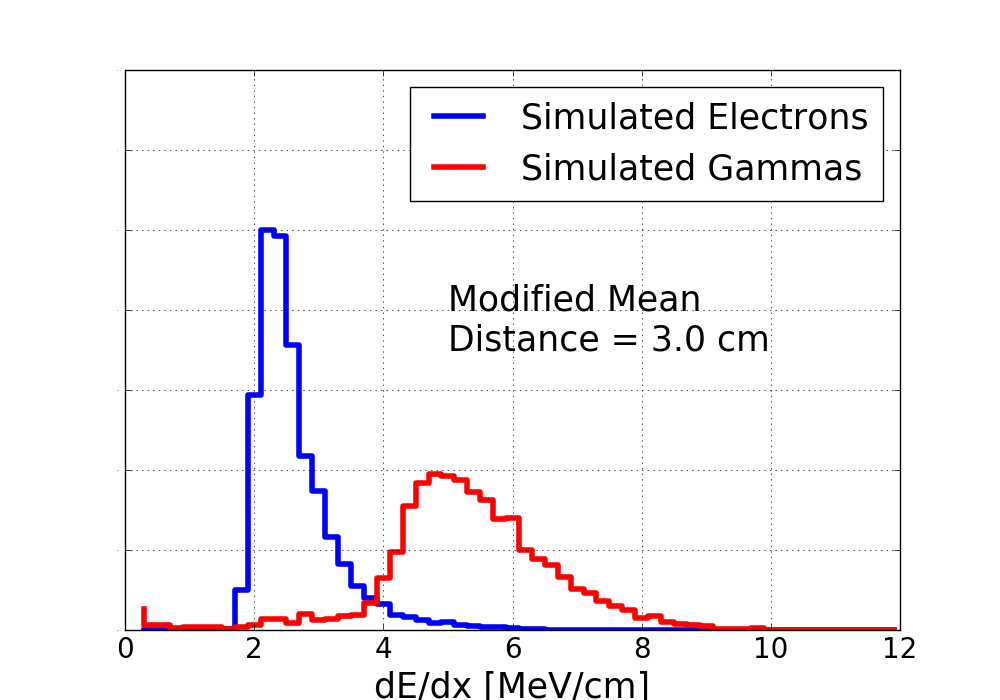}
  \includegraphics[width=0.32\textwidth]{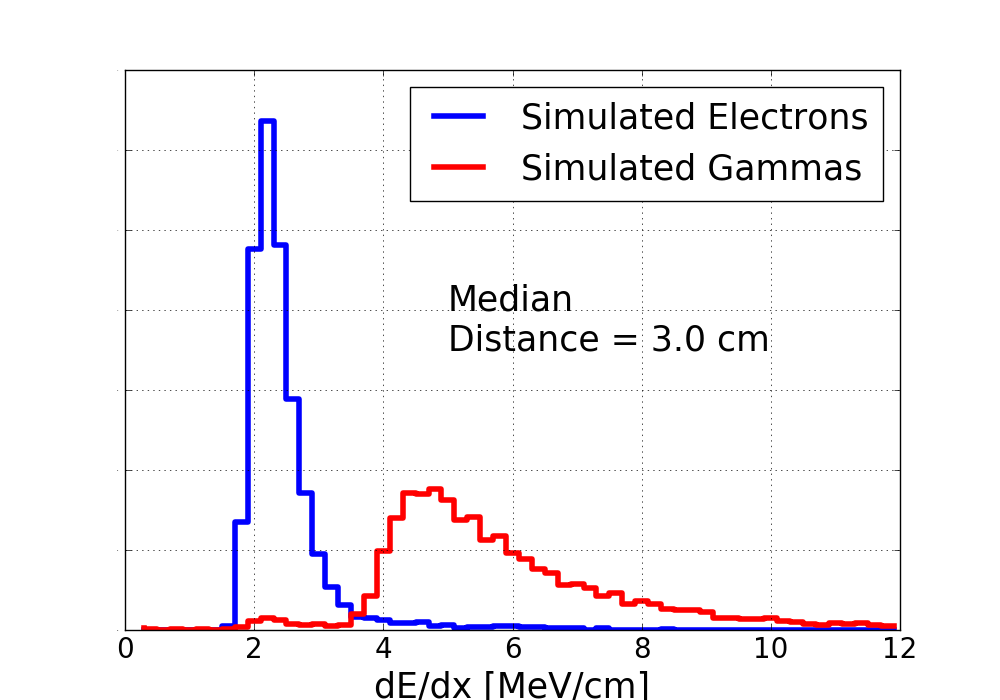}
  \includegraphics[width=0.32\textwidth]{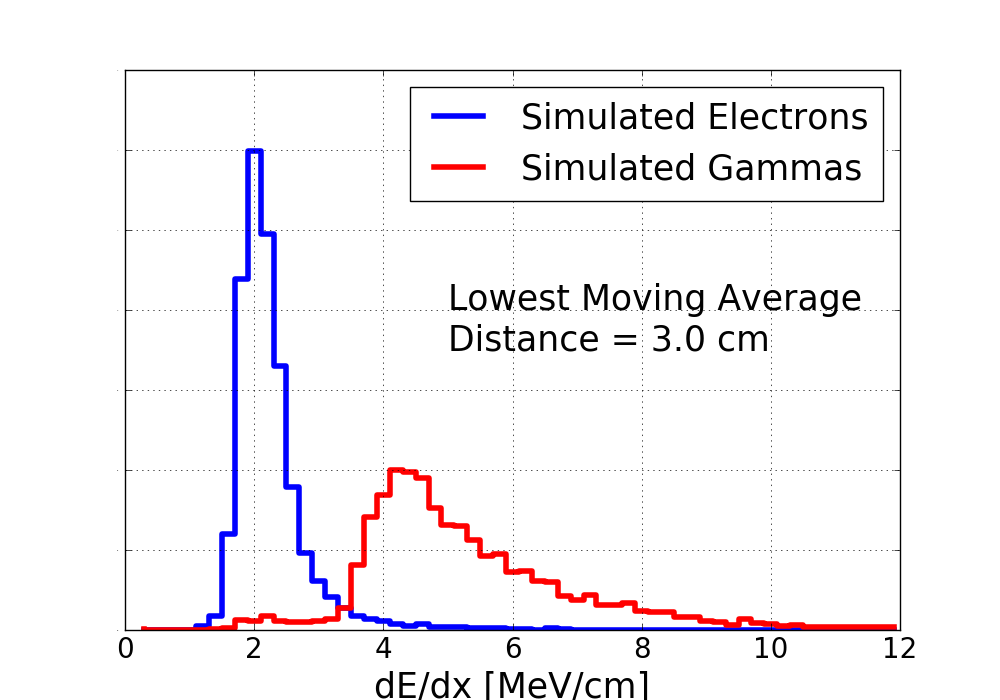}
  
  \includegraphics[width=0.32\textwidth]{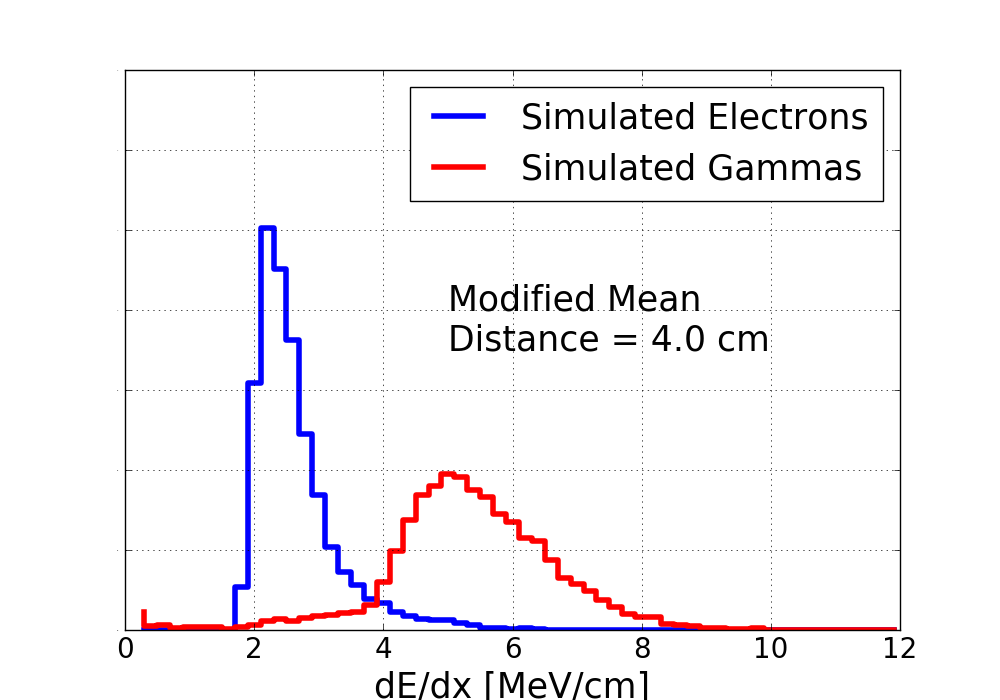}
  \includegraphics[width=0.32\textwidth]{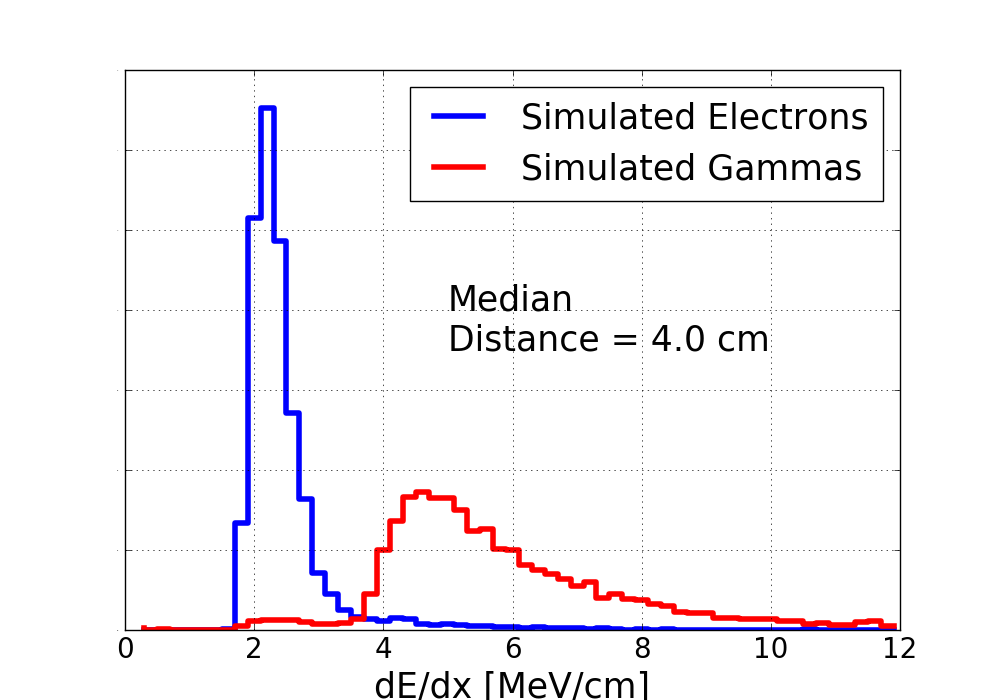}
  \includegraphics[width=0.32\textwidth]{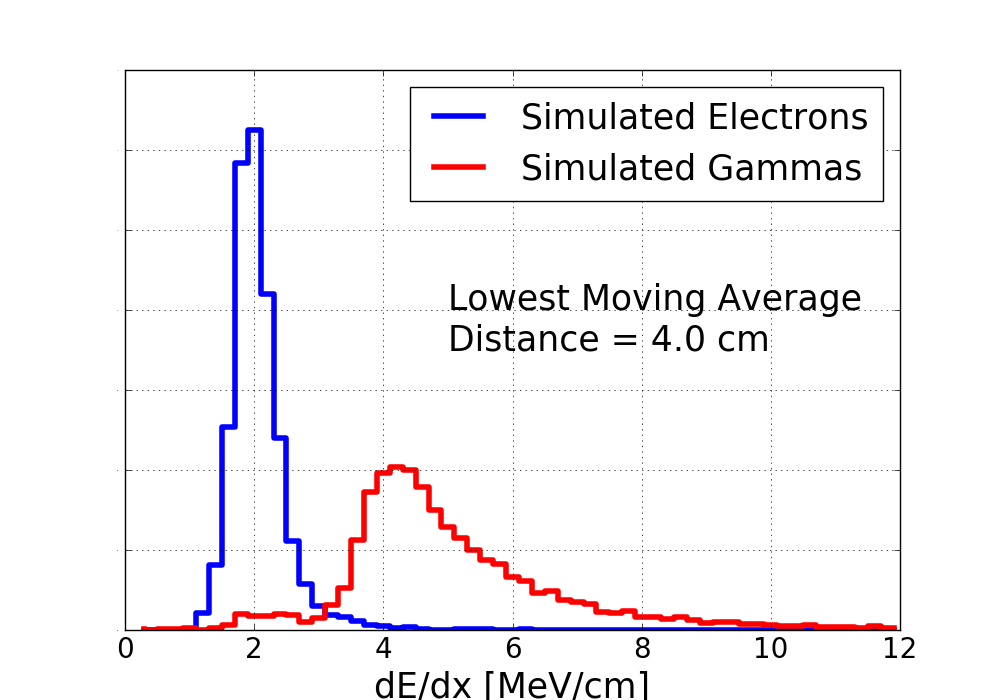}
  
  \includegraphics[width=0.32\textwidth]{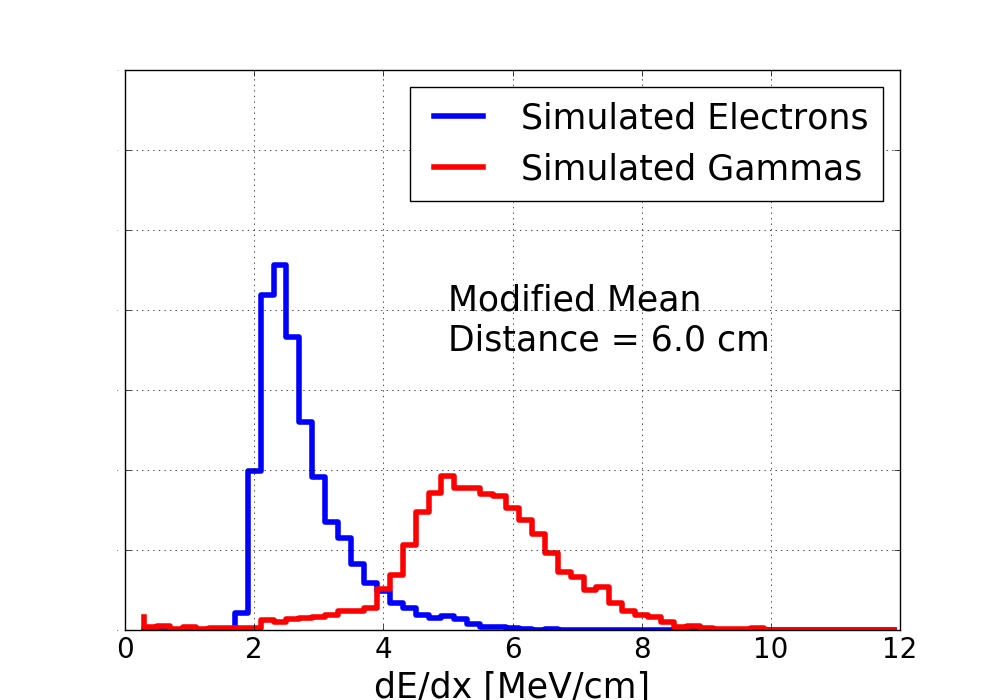}
  \includegraphics[width=0.32\textwidth]{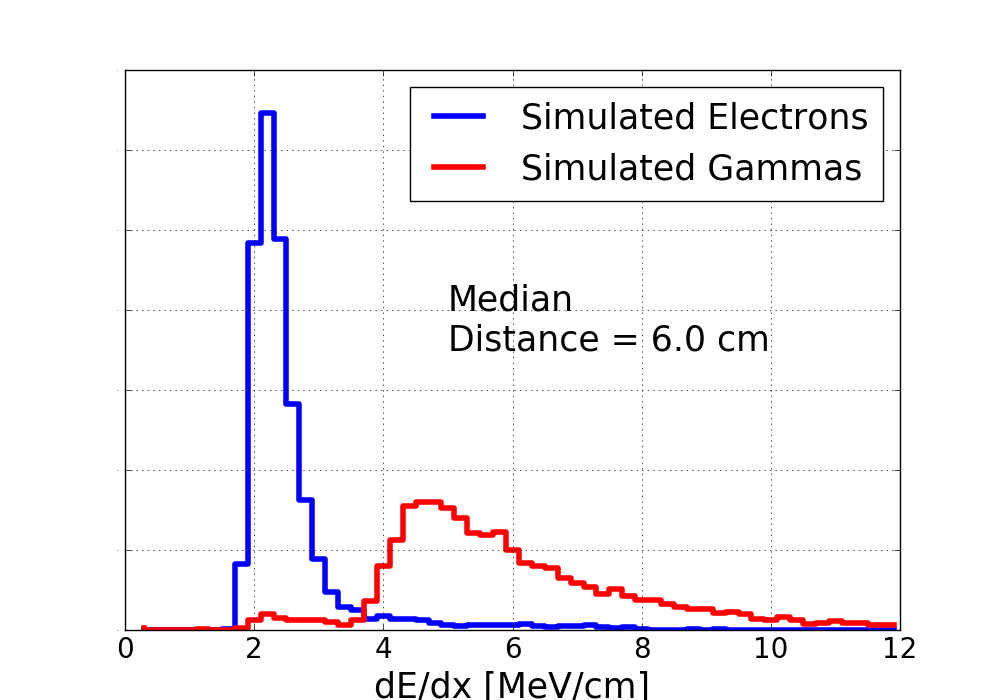}
  \includegraphics[width=0.32\textwidth]{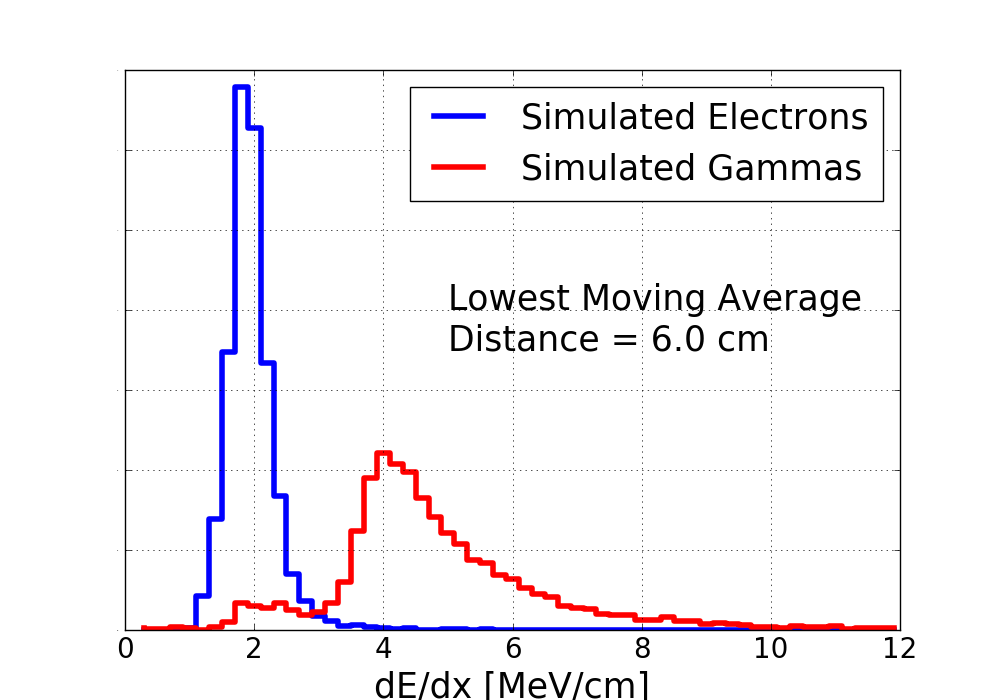}
  
  \includegraphics[width=0.32\textwidth]{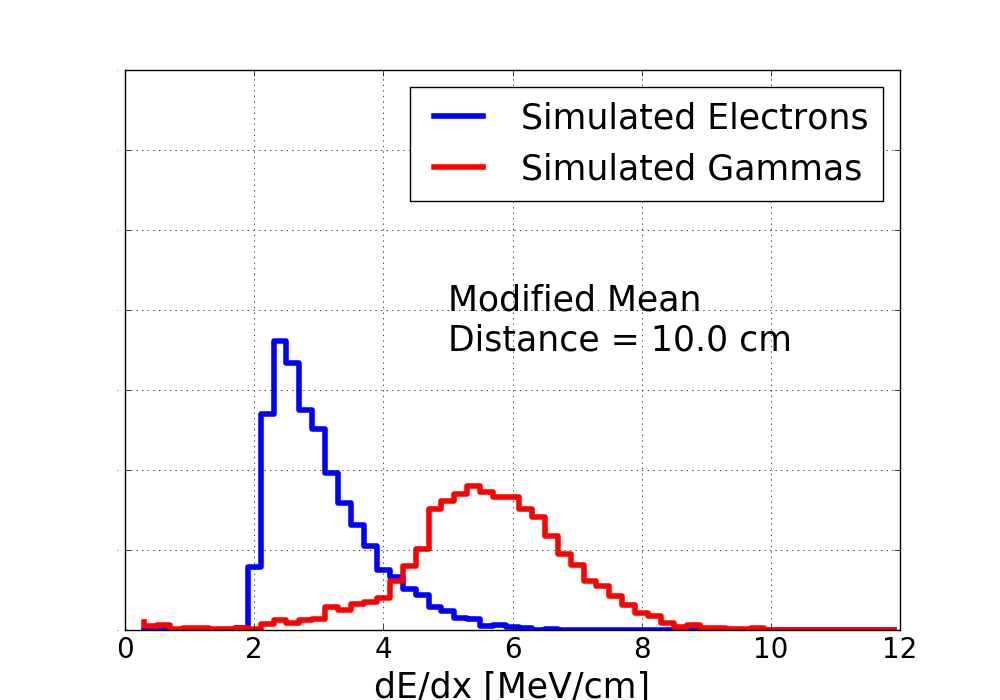}
  \includegraphics[width=0.32\textwidth]{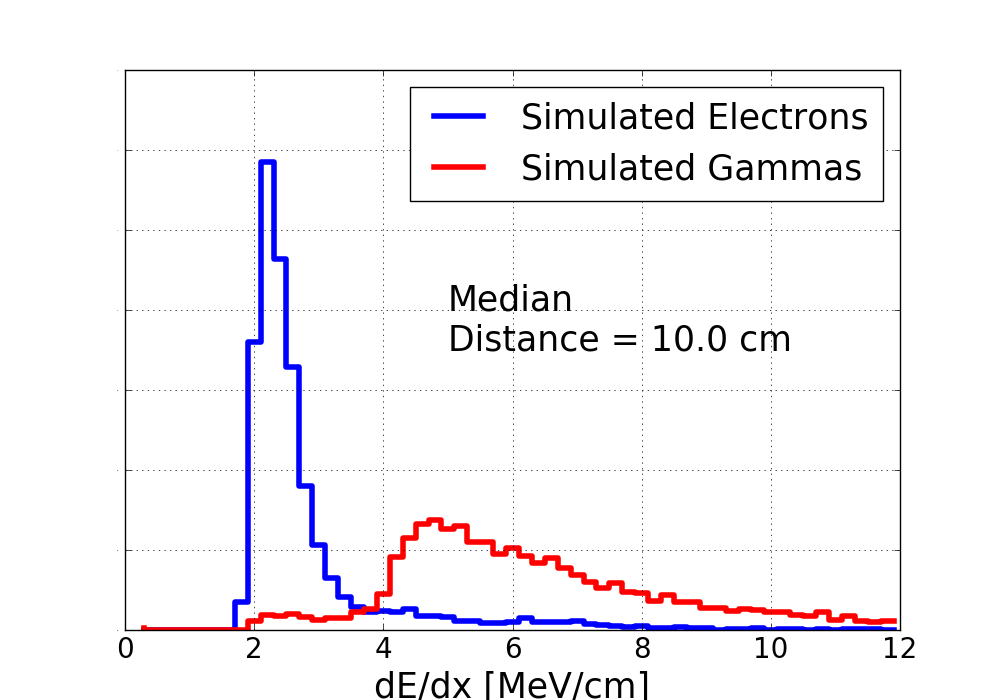}
  \includegraphics[width=0.32\textwidth]{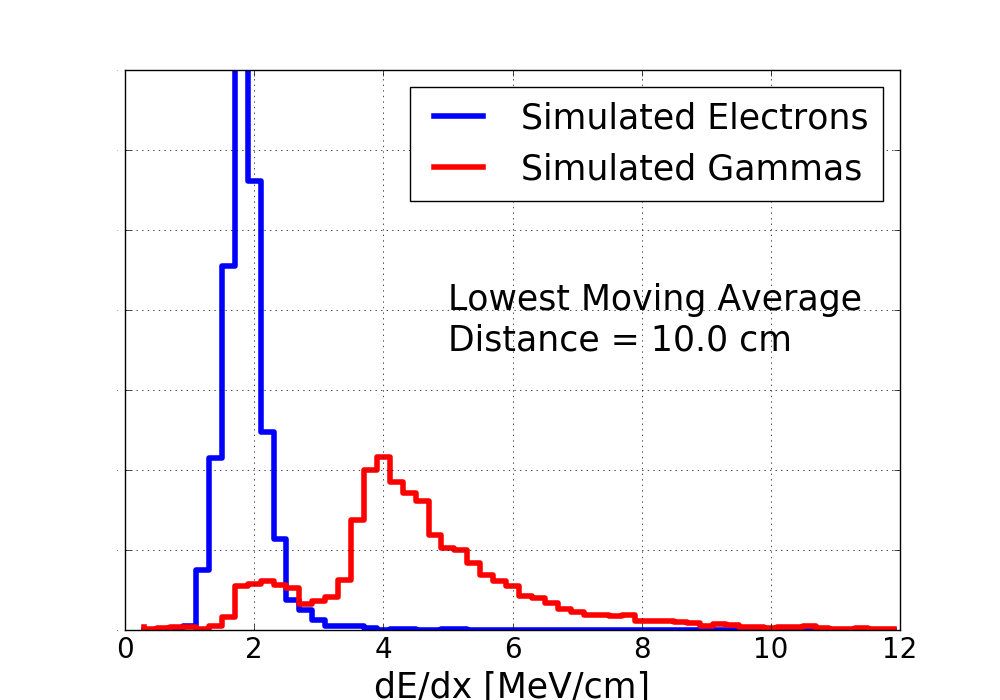}
  \caption{The separation power of the three $dE/dx$ metrics, using a variable amount of the start of the shower in the calculation.  As can be seen, all three metrics show promise at shortest distances.  However, at long distances, the Modified Mean develops a large tail in the electron distribution, and the Lowest Moving Average shifts many gammas into the electron peak.}
  \label{fig:dedx_metrics}
\end{figure*}

To determine which metric is the best for separating electrons from gammas, the truth level energy depositions from the Monte Carlo simulation are examined.  For each event, the true energy depositions are binned into ``hits'' with a pitch that corresponds to the pitch of the simulated shower on the collection plane.  Then, the three $dE/dx$ metrics above are computed for the true hits, and this processes is repeated while varying the length of the shower used in the $dE/dx$ calculation.  The number of hits used in the calculation is a function of the distance along the shower, from the start and moving along the axis of the shower, from which the hits are collected.  The distance used is varied from 2 cm up to 20 cm, with a width of 1 cm.  It was found that a width 1 cm was sufficient to collect the hits along the trunk of the shower.  The results are provided in Figure~\ref{fig:dedx_metrics}, which show that the median metric is the most robust over a variety of distances used at the start of the shower.  Give this result, the median is chosen as the optimal metric for this paper.

In addition, the length of the shower used in this analysis is fixed at 4 cm.  As shown in Figure~\ref{fig:dedx_metrics}, even the median metric begins to degrade at longer distances along the shower, though the degradation is much slower than with the other two methods.  The exact distance used is not the most important parameter.  Between 3 and 5 cm of distance along the shower, all distances yield equivalent separation power.

\begin{figure*}[htb]
  \centering
  \includegraphics[width=0.95\columnwidth]{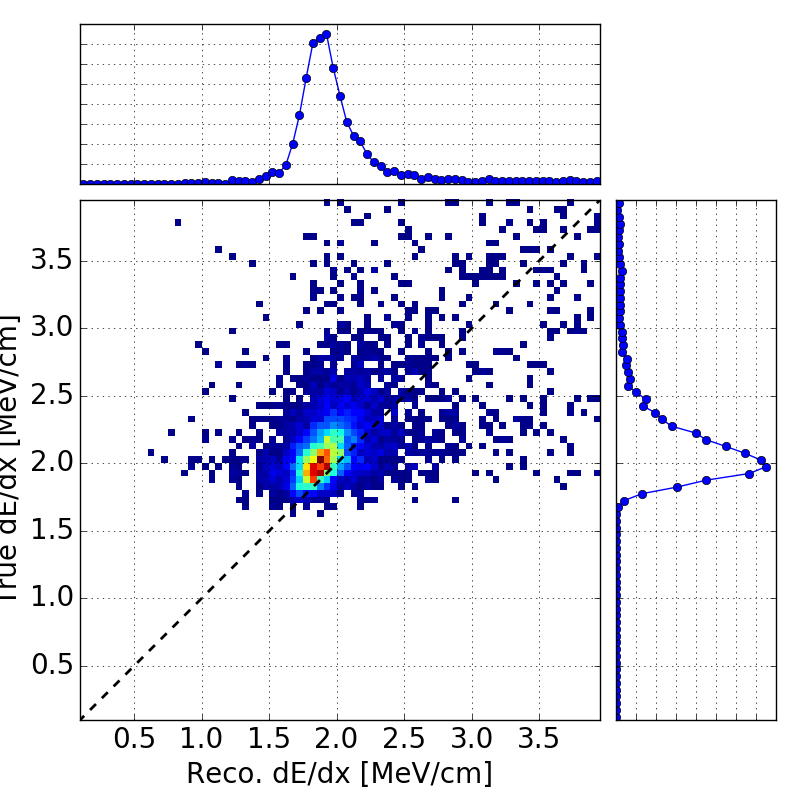}
  \includegraphics[width=0.95\columnwidth]{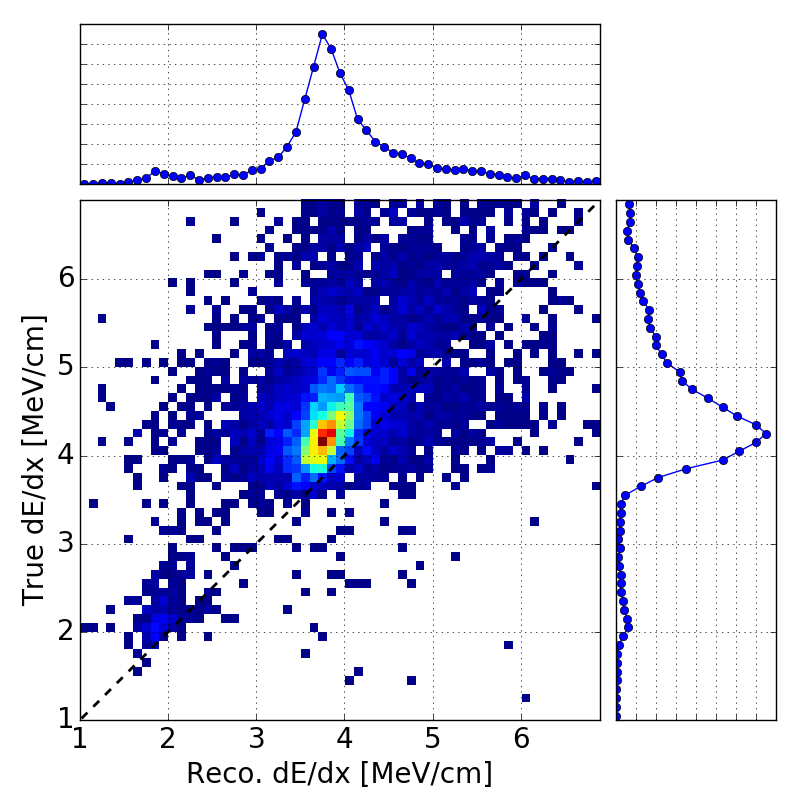}
  \caption{The true $dE/dx$ of the beginning of simulated showers, calculated from simulated energy depositions in the TPC, vs the reconstructed $dE/dx$ of the same showers.  The electrons (left) and the gammas (right) both show a strong correlation between true and reconstructed $dE/dx$. There is a small offset arising from reconstruction inefficiencies, below the 10\% level in both electrons and gammas.}
  \label{fig:true_dedx}
\end{figure*}

Lastly, to verify that the $dE/dx$ calculation from the reconstruction accurately models the true $dE/dx$ of the electromagnetic showers, Figure~\ref{fig:true_dedx} shows the relationship between the true $dE/dx$ and the reconstructed $dE/dx$.  This demonstrates that the reconstructed $dE/dx$ well reproduces the true $dE/dx$ of each shower.

\bibliography{bibliography}

\end{document}